\documentclass[conference]{IEEEtran}
\IEEEoverridecommandlockouts
\usepackage{cite}
\usepackage{amsmath,amssymb,amsfonts}
\usepackage{algorithmic}
\usepackage{graphicx}
\usepackage{textcomp}
\usepackage{xcolor}
\usepackage[hyphens]{url}

\usepackage{formatting/packages}
\usepackage{formatting/macros}

\def\BibTeX{{\rm B\kern-.05em{\sc i\kern-.025em b}\kern-.08em
    T\kern-.1667em\lower.7ex\hbox{E}\kern-.125emX}}

\title{\textsc{DaCapo}: Accelerating  Continuous Learning in Autonomous Systems for Video Analytics}

\author{
\IEEEauthorblockN{
    Yoonsung Kim\IEEEauthorrefmark{0}\customspace
    Changhun Oh\IEEEauthorrefmark{0}\customspace
    Jinwoo Hwang\IEEEauthorrefmark{0}\customspace
    Wonung Kim\IEEEauthorrefmark{0}\customspace
    Seongryong Oh\IEEEauthorrefmark{0}\\
    Yubin Lee\IEEEauthorrefmark{0}\customspace
    Hardik Sharma\IEEEauthorrefmark{2}\IEEEauthorrefmark{1}\customspace
    Amir Yazdanbakhsh\IEEEauthorrefmark{3}\customspace
    Jongse Park\IEEEauthorrefmark{0}
}
\vspace{1ex}
\IEEEauthorblockA{
    \IEEEauthorrefmark{0}KAIST\customspace
    \IEEEauthorrefmark{2}Meta\customspace
    \IEEEauthorrefmark{3}Google DeepMind\\
}
\vspace{1ex}
\IEEEauthorblockA{
    \bluetext{\{yskim, choh, jwhwang, wukim, sroh, yblee\}@casys.kaist.ac.kr}\\
    \bluetext{hardiksharma@meta.com}\customspace
    \bluetext{ayazdan@google.com}\customspace
    \bluetext{jspark@casys.kaist.ac.kr}
}
\thanks{\IEEEauthorrefmark{1}This work was done at Google.}
}

\begin{document}
\maketitle
\thispagestyle{plain}
\pagestyle{plain}


\begin{abstract}
Deep neural network (DNN) video analytics is crucial for autonomous systems such as self-driving vehicles, unmanned aerial vehicles (UAVs), and security robots.
However, real-world deployment faces challenges due to their limited computational resources and battery power.
To tackle these challenges, continuous learning exploits a lightweight ``student'' model at deployment (\textbf{\underline{\smash{\textit{inference}}}}), leverages a larger ``teacher'' model for labeling sampled data (\textbf{\underline{\smash{\textit{labeling}}}}), and continuously retrains the student model to adapt to changing scenarios (\textbf{\underline{\smash{\textit{retraining}}}}). 
This paper highlights the limitations in state-of-the-art continuous learning systems: (1) they focus on computations for retraining, while overlooking the compute needs for inference and labeling, (2) they rely on power-hungry GPUs, unsuitable for battery-operated autonomous systems, and (3) they are located on a remote centralized server, intended for multi-tenant scenarios, again unsuitable for autonomous systems due to privacy, network availability, and latency concerns. 
We propose a hardware-algorithm co-designed solution for continuous learning, \dacapo, that enables autonomous systems to perform concurrent executions of inference, labeling, and retraining in a performant and energy-efficient manner.  
\dacapo comprises (1) a spatially-partitionable and precision-flexible accelerator enabling parallel execution of kernels on sub-accelerators at their respective precisions, and (2) a spatiotemporal resource allocation algorithm that strategically navigates the resource-accuracy tradeoff space, facilitating optimal decisions for resource allocation to achieve maximal accuracy.
Our evaluation shows that \dacapo achieves 6.5\% and 5.5\% higher accuracy than a state-of-the-art GPU-based continuous learning systems, Ekya and EOMU, respectively, while consuming 254$\times$ less power.
\end{abstract}
\section{Introduction}
\label{sec:intro}
Live video analytics is key to enable seamless functionality of interactive autonomous systems, such as autonomous driving vehicles~\cite{iccv-2023-gameformer, haltingproblem-chi-2023, cvpr-2023-xiong, cvpr-2023-jia, st-p3, TBP-Former, Tijtgat}, unmanned aerial vehicles (UAVs)~\cite{iccv-2023-aerialvln, iccv-2023-li, GLTF-MA, UAVMOT, VTUAV}, and security surveillance robots~\cite{icra-2023-pichierri, AU-Air, ral-2022-seo, 8794073, icra-2021-pheromone, ral-2021-gridnet, 9562043}. 
Real-time analysis of live videos is indispensable for comprehending rapidly changing environmental conditions, a pivotal aspect that ensures the successful execution of the autonomous systems' missions.

\begin{figure}[t]
        \centering
        \includegraphics[width=1.0\linewidth]{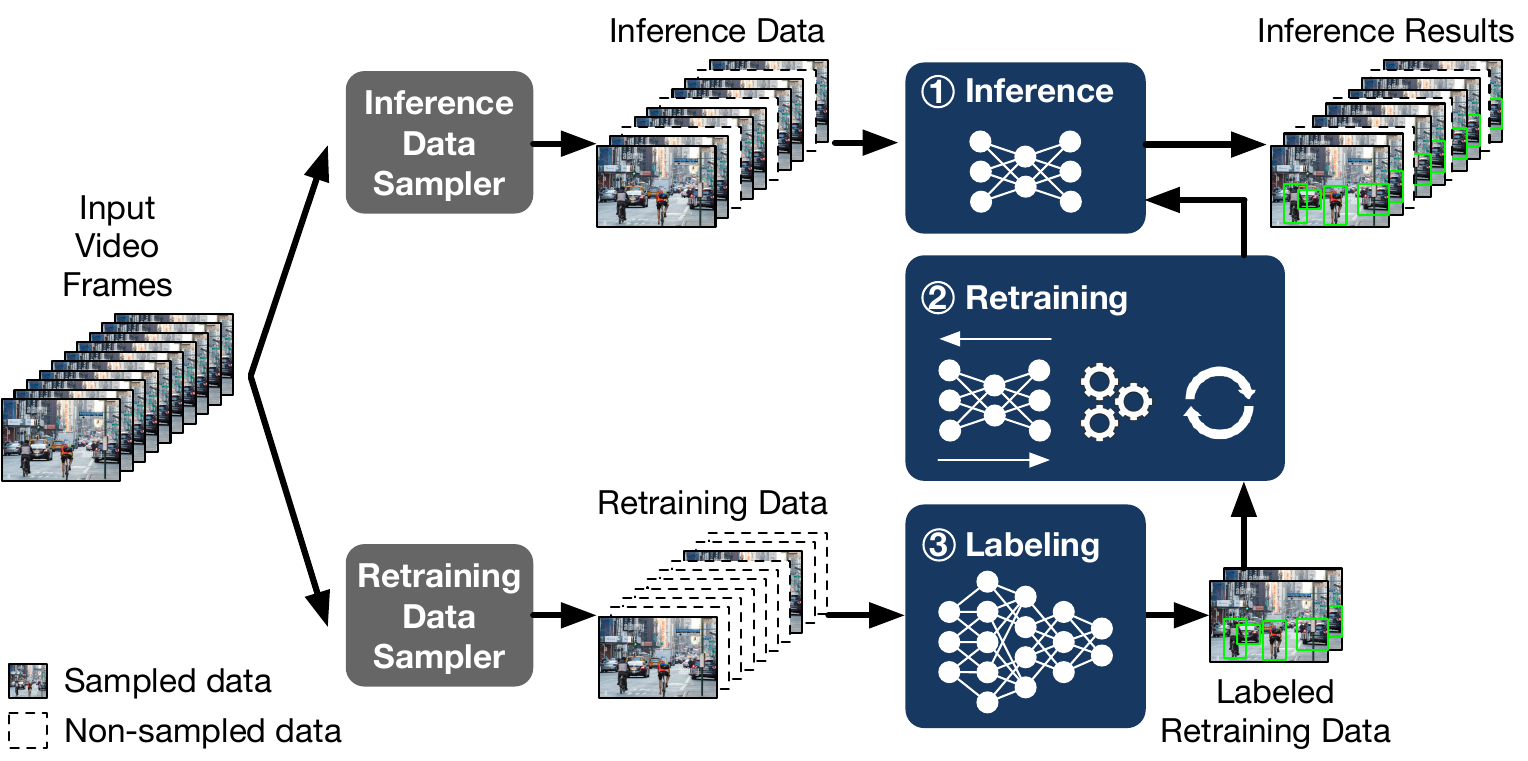}
        \caption{Overview of continuously learning video analytics on autonomous systems. To address privacy, networking cost, and latency concerns, autonomous systems exclusively use constrained computing resources to concurrently execute the three continuous learning kernels -- (1) inference, (2) retraining, and (3) labeling -- which presents a performance challenge.}
        \vspace{-3ex}
        \label{fig:continual-learning}
\end{figure}

Modern video analytics heavily relies on the capabilities of deep neural networks (DNNs). 
However, the computational demand to execute DNN inference on individual video frames is far beyond the capabilities of real-world systems.
Although sampling can decrease computational demands, it often compromises accuracy, especially in videos captured by moving cameras, a frequent condition in autonomous systems.

To address this issue, \textit{continuous learning} (CL), an emerging algorithmic methodology, has gained significant attention in recent years~\cite{hpca-2024-usas, eomu-mm-2023, recl-nsdi-2023, ekya-nsdi-2022, ghunaim-cl-23-cvpr, online-distillation}.
This innovative approach enables autonomous systems to perform inference using lightweight ``student'' DNN models, which results in reduced resource consumption.
However, the limited generality of the lightweight student models gives rise to the \textit{data drift} problem, where live video data deviates from training data.
Continuous learning incorporates constant \textit{on-the-spot} retrainings, which introduce a new computational challenge, as the systems must now handle not only \textit{inference}, but also \textit{labeling} and \textit{retraining}, as illustrated in Figure~\ref{fig:continual-learning}.
We denote these three tasks as \textit{continuous learning kernels} in this paper.
Two recent pioneering works, Ekya~\cite{ekya-nsdi-2022} and RECL~\cite{recl-nsdi-2023}, propose initial system designs for continuously learning video analytics.
Although these works have paved the way for a new research direction, their direct deployment to autonomous systems is infeasible due to the following reasons: (1) these systems overlook the necessity of computation for labeling and inference, a factor that could be substantial for battery-operated, resource-constrained autonomous systems; (2) they assume excessively rich computation resources for autonomous systems; and (3) they are designed to be deployed on a remote centralized server, intended for multi-tenant scenarios.
For instance, Ekya~\cite{ekya-nsdi-2022} runs on a multi-GPU edge server and RECL~\cite{recl-nsdi-2023} offloads the retraining computation to the scalable cloud, both largely focusing on the retraining computations of multiple models. 
In this paper, we propose a hardware-software co-designed acceleration solution, \dacapo, that overcomes the limitations of existing systems and enables effective deployment of continuously learning video analytics on autonomous systems without the need for remote access to abundant resources.
In designing \dacapo, we identify the following challenges:
\begin{itemize}[labelindent=0.5em,nolistsep,leftmargin=2.0em]
\item \textbf{Challenge 1:} \dacapo must offer significantly higher performance and energy efficiency compared to the existing high-performance computing platforms such as GPUs, to effectively implement continuous learning for video analytics on battery-operated autonomous systems.
\item \textbf{Challenge 2:} \dacapo requires a flexible and dynamically reconfigurable accelerator architecture to rapidly adapt to the changing resource demands of the three continuous learning kernels when data drifts occur.
\item \textbf{Challenge 3:} \dacapo must be able to capture the resource-accuracy tradeoff of the continuous learning kernels and efficiently utilize the computing resources to maximize the resulting accuracy.
\end{itemize}

To address the aforementioned challenges, we make the following contributions.
\begin{itemize}[labelindent=0.5em,nolistsep,leftmargin=2.0em]
\item \textbf{Spatially-partitionable and precision-flexible accelerator architecture. } We architect the \dacapo accelerator, employing a spatially-partitionable architecture, which can be dynamically composed into sub-accelerators (SAs).
The design principle aims to attain flexibility for better resource utilization without imposing unnecessarily large overhead to enable this adaptability.
To this end, the \dacapo accelerator is organized as vertically-stacked rows of processing elements in which the rows can be partitioned and grouped into two sub-accelerators.
A processing element in \dacapo performs a vector dot product rather than a multiplication, which we call Dot-Product Engine (DPE), leveraging recently proposed MX-based precision-flexible arithmetic to unleash higher performance and energy-efficiency from the accelerator. 
This spatial sharing and precision configurability are the keys that enable \dacapo to obtain orders of magnitude higher efficiency and allow the three kernels to fully utilize the given resources for maximal accuracy.

\item \textbf{Spatiotemporal resource allocation algorithm for maximal accuracy.} 
Resource requirements for the three kernels vary depending on the runtime scenarios that constantly change. 
A key insight driving our resource allocation algorithm is that \emph{labeling} plays a pivotal role in the occurrence of data drifts for promptly collecting retraining data with new data patterns, requiring relatively more computing resources.
To maximize accuracy using the limited resources, we devise a spatiotemporal resource allocation algorithm that makes static (spatial) and dynamic (temporal) decisions for resource allocations.
First, at offline, the algorithm statically partitions the accelerator into two sub-accelerators, T-SA and B-SA, based on the resource requirement of inference to keep up with the input frame rate.
Then, during runtime, the algorithm dynamically detects data drifts through constant accuracy validations 
and when it is detected, the algorithm allocates more time slots for labeling until 
the dataset is sufficiently updated (temporal decision). 
\end{itemize}
To evaluate \dacapo's effectiveness, we use three pairs of (student, teacher) object recognition models (see Table~\ref{tab:benchmark-model}) and the BDD100K driving video dataset.
We model three types of data drift: (1) task change by addition or removal of object labels, (2) transitions between daytime and nighttime, and (3) location change between city and highway. 
These data drifts are mixed to synthesize different workload scenarios representing diverse real-world situations that autonomous systems would face. 
We develop \dacapo using Verilog RTL and synthesize it using Synopsys Design Compiler with TSMC 28nm technology.
To cross-validate the cycle-accurate behaviors of \dacapo, we also develop an in-house simulator built upon SCALE-Sim~\cite{scalesim}.
Our evaluation shows that \dacapo achieves 6.5\% and 5.5\% higher accuracy than the two state-of-the-art GPU-based continuous learning systems, Ekya and EOMU, respectively.
Furthermore, we observe that \dacapo achieves the accuracy improvement, consuming 254$\times$ less energy compared to the GPU baseline.
These advantages highlight \dacapo's role in connecting the algorithmic advancements in continuous learning to their practical deployment on autonomous systems.
Our tools, including system and hardware simulators and a continuous learning benchmark suite with datasets, are available at \bluetext{\url{https://github.com/casys-kaist/DaCapo}}. 
\section{Background}
\label{sec:background}

This paper aims to develop an acceleration solution for continuously learning video analytics on autonomous systems. 
In this section, we briefly outline the target problem and introduce existing solutions.
\subsection{Video Analytics in Autonomous Systems}
Modern autonomous systems leverage compute-heavy deep neural network (DNN) models for complex video analytics. 
Constrained by limited resources, these systems struggle to meet latency requirements, often requiring compromises that can entail frame drops and a potential reduction in overall analytical quality. 
Recently, using less capable yet lightweight DNN models has been introduced as a viable solution~\cite{neuos, zerobn, collate, 10247965}, allowing the systems to match the inference rate with the frame rate. 
However, the customized models present a new challenge known as \textit{data drift}, where live video data diverges from the training data, consequently reducing the accuracy of the lightweight models. 
\subsection{Continuously Learning for Video Analytics at Edge}
To address the data drift problem, a large body of literature~\cite{hpca-2024-usas, recl-nsdi-2023, active-cl-sensys-2022, eomu-mm-2023, odin, ams, ekya-nsdi-2022} has explored \textit{continuous learning}, which aims to continuously adapt models to newly appearing data.
This approach allows the models to stay up-to-date with the changing data distribution.
While continuous learning is a promising approach to reduce compute demand for inference, it introduces new compute load for constant retraining over newly collected data as well as labeling the retraining data by running inference for large, state-of-the-art models, often referred to as \textit{teacher} models~\cite{omd}.
The challenge arises from the necessity of executing the three distinct yet compute-intensive kernels, each affecting the final accuracy differently, creating a complicated tradeoff space.

\subsection{Continuous Learning Systems for Video Analytics} 
Inspiring prior works, Ekya~\cite{ekya-nsdi-2022}, and RECL~\cite{recl-nsdi-2023}, are pioneering efforts that address the challenge, approaching it from a system design perspective. 
Ekya and RECL offer resource allocation and scheduling solutions tailored to GPU-accelerated live video analytics platforms.
In particular, they address the ``at-scale'' problem, where the systems are required to manage and process multiple DNN models specific to various video streams concurrently.
Optimizing for multi-tenant workload scenarios, the existing systems are centered on facilitating resource sharing across different models, requiring the use of mid-size to cloud-scale servers equipped with multiple compute-powerful computing platforms such as datacenter-level GPUs.
A recent work, EOMU~\cite{eomu-mm-2023}, takes a similar yet different approach by proposing a log-based adaptive resource allocation technique for edge-cloud continuous learning system. 
In contrast, this work aims to focus on enabling single-task video analytics on resource-constrained autonomous systems running the three continuous learning kernels, a distinctive and novel challenge.
\section{Challenges and Opportunities}
\label{sec:motiv}
In this section, we examine the limitations of current solutions, identify key challenges, and explore potential opportunities for improvement. 

\subsection{Unveiling the Dilemma in Continuous Learning}
\label{sec:feasibility}
To understand the performance implications, we perform a preliminary experimental study, using two pairs of \underline{S}tudent and \underline{T}eacher object recognition models: (\underline{S}:ResNet18, \underline{T}:WideResNet50) and (\underline{S}:ResNet34, \underline{T}:WideResNet101).
For the experiments, we use an NVIDIA RTX 3090 and an NVIDIA Jetson Orin, representing datacenter-scale and autonomous system-level GPUs respectively.
As a baseline for continuous learning systems, we built Ekya under an idealized assumption that the system can optimally utilize the provided GPU resources to obtain the maximum achievable accuracy. 

\begin{figure}[t]
    \centering
    \includegraphics[width=\linewidth]{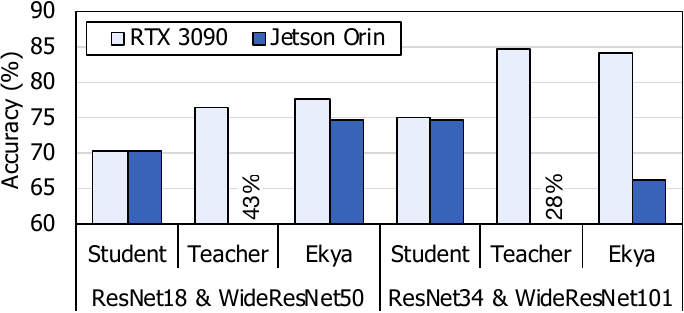}
    \caption{Accuracy comparisons of Ekya versus student and teacher models. Student and teacher models are \emph{non}-continuous learning cases. The experiments are conducted on RTX 3090 and Jetson Orin. The accuracy gap between the two GPUs is attributed to inevitable frame drops due to a lack of computing resources.}
    \vspace{-2ex}
    \label{fig:feasibility-analysis}
\end{figure}

Figure~\ref{fig:feasibility-analysis} reports the accuracy results of student/teacher models compared to the Ekya baseline, measured by calculating the ratio of correct predictions to the total number of classifications.
Dropped frames produced by insufficient computing resources are considered as incorrect results.
The RTX 3090 has sufficient compute resources, resulting in no frame drops.
Therefore, the results are solely evaluated by their algorithmic performance. 
The teacher models always substantially outperform student models, exhibiting their algorithmic superiority. 
While Ekya employs the student model for inference, its accuracy often approaches or even exceeds that of the teacher model since Ekya can adeptly customize the student model to suit the specific system environment.
However, as we change the compute platform from RTX 3090 to Orin, for both teacher and Ekya baselines we observe significant accuracy drops. 
These results demonstrate that continuous learning effectively mitigates computational bottlenecks, but still notable accuracy loss persists when insufficient resources are provided, the key challenge addressed in this work.
\textit{Existing continuous learning systems function effectively only with substantial computational resources, a condition incompatible with the limitations of autonomous systems. 
This insight underscores the necessity for specialized hardware innovation, crucial for obtaining levels of performance and efficiency otherwise unachievable, thereby unleashing the full potential of continuous learning in autonomous systems.}

\subsection{Workload Characterization of the Three Kernels}
\begin{figure}[t]
    \centering
    \includegraphics[width=\linewidth]{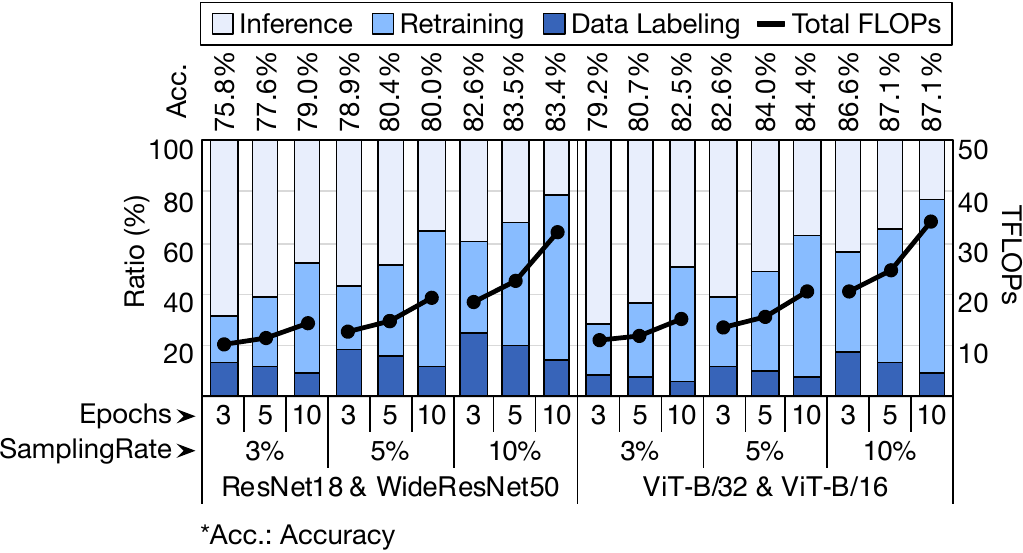}
    \caption{MAC operation breakdown of the three kernels and total FLOPS for the entire experiment runs.}
    \vspace{-2ex}
    \label{fig:operation-breakdown}
\end{figure}

A straightforward solution for the aforementioned challenge is to develop a high-performance accelerator with ample compute resources and exploit time-multiplexing to flexibly schedule the three kernels interchangeably on the chip.
However, such a time-sharing technique would limit resource utilization since the three kernels offer different levels of parallelism (e.g., inference offers much lower data parallelism than retraining).
Alternatively, the three kernels can be spatially parallelized by designing an accelerator with three separate sub-modules, with each module dedicated to a specific kernel.
We found that this approach is also suboptimal since the computational demands of the kernels exhibit temporal variations at runtime. 
To understand the workload of continuous learning, we explore the computational demands of running a continuous learning system for 120 seconds, measuring these in FLOPs, and examine how each of the three kernels contributes to the overall FLOPs count.
Figure~\ref{fig:operation-breakdown} presents the results, with the stacked bars illustrating the proportional distribution of FLOPS for inference, retraining, and labeling from top to bottom, while the line graph delineates how the total FLOPs scale over time.
We observe a notable shift in proportions: the proportion allocated to retraining surges from 26.0\% to 82.3\% as data sampling rates for labeling and number of epochs for retraining increase, while inference and labeling decrease from 57.8\% and 27.1\% to 9.1\% and 7.0\% respectively.
The results also show that increasing the sampling rate and number of epochs improves accuracy, albeit at the expense of increased FLOPs.
\textit{The overall accuracy of a continuous learning system is jointly influenced by the interplay of the three kernels, whose contributions shift based on hyperparameters, leading to varying computational demands. Inspired by this insight, we design the \dacapo accelerator to enable dynamic and flexible resource allocation among the kernels. This approach aims to optimize resource efficiency, consequently improving the final accuracy.}
\subsection{Opportunities from Low-Precision Arithmetics}

As discussed earlier, effective continuous learning in an autonomous system demands not just exceptional performance but also significant energy efficiency, which motivates the exploration of strategies that can achieve enhanced performance and efficiency, without sacrificing accuracy.
Low-precision arithmetic through quantization has been widely adopted to significantly reduce the computational resource demands for training and inference~\cite{hbfp:2018, fast:2022, flexpoint:2017, mx-isca-2023}.
Among various quantization formats, block floating point (BFP) has recently gained prominence owing to its hardware-friendly characteristics and ability to support a wide range of real values~\cite{hbfp:2018, fast:2022, mx-isca-2023}. 
BFP groups a set of floating point values, forces them to have a shared exponent by shifting the mantissa accordingly, and stores the group of truncated mantissa bits along with the shared exponent.
This way, most computations happen in the integer domain, which is significantly cheaper than single-precision floating-point (FP32) arithmetics, while still offering FP32-\emph{like} algorithmic behaviors.
Particularly, a recent work from Microsoft~\cite{mx-isca-2023} proposes a customized block floating point format, called MX (Micro-eXponent), and demonstrates the effectiveness on both training and inference. 
The prior work also suggests that in general, training requires higher precision (MX9) than inference (MX6 or MX4), and the best-fit MX precision varies depending on the particular models.
However, the MX accelerator is not precision-flexible, lacking support for multiple precisions on a single hardware, which is not well-suited for continuous learning scenarios where an accelerator should simultaneously perform both training and inference for different models.
Therefore, to effectively employ MX and fully realize the potential for continuous learning, there is a need for customized architectural techniques, which is one of the goals of this work.  
\textit{While MX offers promising opportunities, the accelerator needs to support various MX precisions, switchable at runtime. This flexibility is essential to fully capitalize on the advantages of mixed computations in both inference and retraining executions across the three kernels.}
\section{Overview of \dacapo's System Workflow}
\label{sec:algorithm}

\begin{figure}[t]
    \centering
    \includegraphics[width=1.0\linewidth]{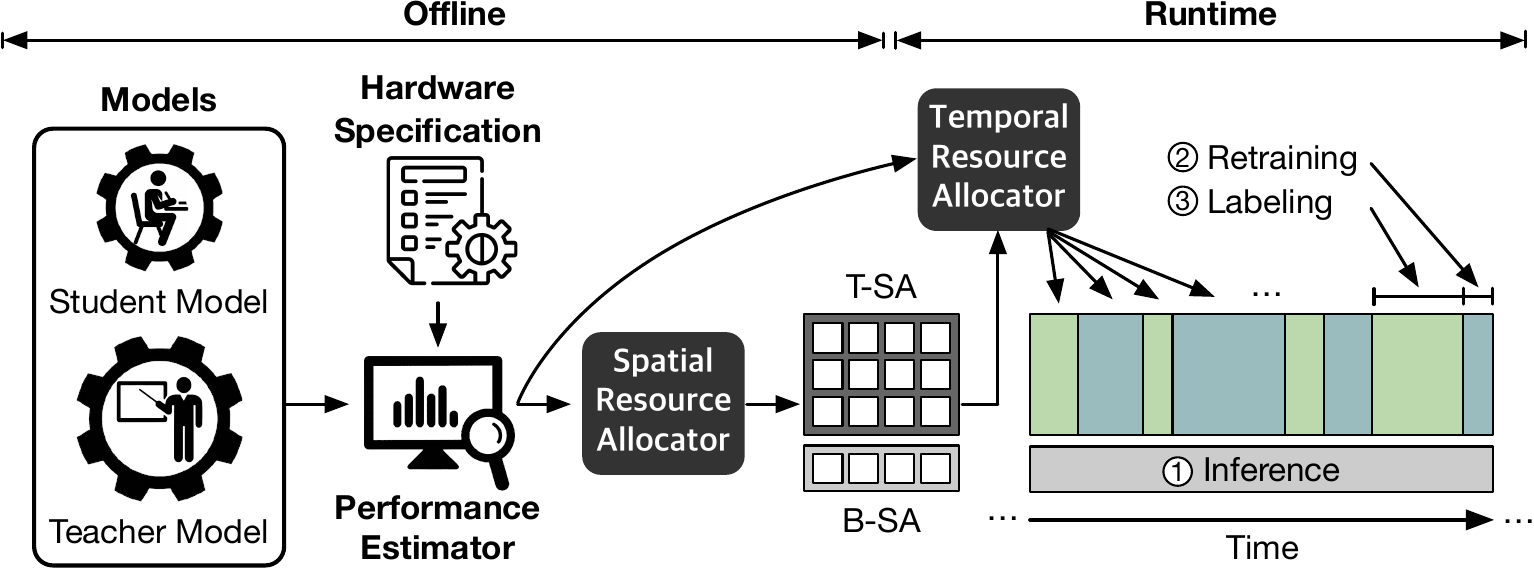}
    \vspace{-2ex}
    \caption{Workflow of a \dacapo-based continuously learning video analytics system.}
    \vspace{-2ex}
    \label{fig:overview}
\end{figure}
Figure~\ref{fig:overview} illustrates the overall workflow of \dacapo-accelerated continuous learning in an autonomous system. 

\niparagraph{\circled{1} System initialization.}
The system needs to be initialized with the student and teacher DNN models for inference and labeling.   
These models are pre-trained over the general dataset without having any specific context that the system is actually used for. 
We assume that the hyperparameters and optimization techniques for on-the-fly retraining are pre-determined and not adjusted at runtime.
Additionally, we also assume that the frame rate of input video and its resolution are a priori known knowledge and never change at runtime.
\niparagraph{\circled{2} Performance estimation.}
The subsequent stage involves our simulation-based performance estimator, which utilizes the architecture of the specified student/teacher models and the hardware specifications of the \dacapo accelerator to approximate the throughput of the three kernels. 
In this phase, the estimator examines every supported MX precision, ranging from MX4 through MX6 to MX9, to assess its impact on accuracy.
Our observations show that the models evaluated achieve stable outcomes when using MX9 for retraining and MX6 for inference/labeling, whereas utilizing MX4 leads to considerable accuracy degradation in both scenarios.
Note that this observation aligns with the experimental findings reported in the original MX paper~\cite{mx-isca-2023}.
This crucial information is then relayed to the resource allocators in subsequent stages, aiding them in their decision-making processes.

\niparagraph{\circled{3} Offline spatial resource allocation.}
Utilizing the performance estimation, our spatial resource allocator calculates the minimum number of rows needed for B-SA to process inferences for input frames, ensuring it matches the input frame rate.
Although the MX precisions remain constant during runtime, precision flexibility is still necessary because a row may be allocated to either T-SA or B-SA based on the decisions made in spatial resource allocation.
\niparagraph{\circled{4} Online temporal resource allocation.}
While B-SA consistently handles inference computations, T-SA alternates between retraining and labeling tasks through time-sharing.
In formulating the temporal resource allocation algorithm, we exploit an empirical insight that when data drifts occur, rapidly labeling data with a new distribution is more crucial than conducting additional retraining iterations.
For \dacapo to respond effectively to data drifts, it must first detect them. 
To achieve this, \dacapo tracks accuracy trends in real-time, employing a small portion of the labeled training dataset as a validation set for identifying any data drifts.
Upon their detection, \dacapo adjusts its time allocation, favoring labeling over retraining. 
This shift in temporal resources aids the system in quickly recovering from accuracy losses caused by data drifts. 
The empirical evidence supporting this approach will be detailed in Section~\ref{sec:experiment}.
\section{\dacapo Accelerator Architecture}
\label{sec:architecture}

\begin{figure}[t]
    \centering
    \includegraphics[width=0.9\linewidth]{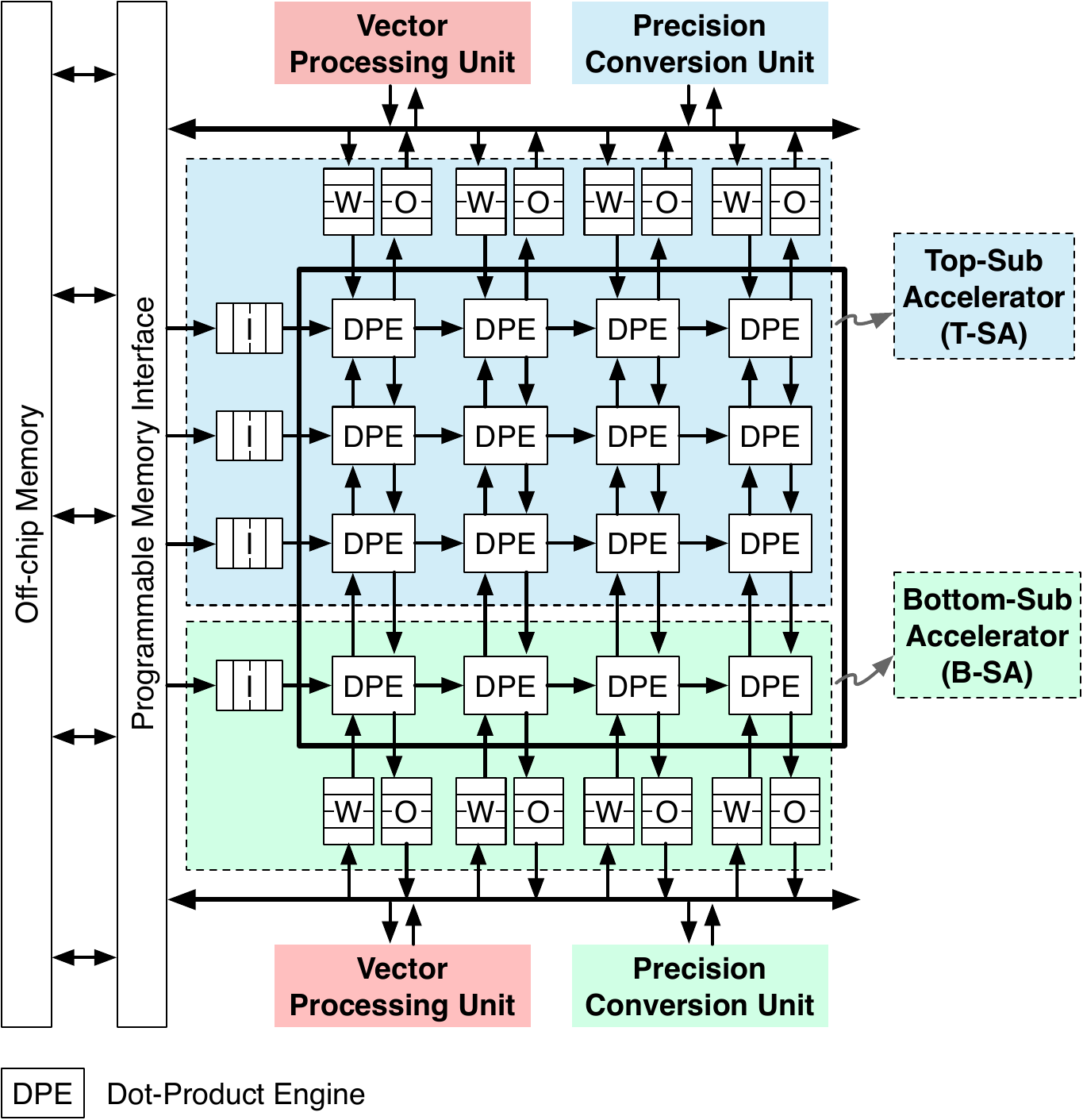}
    \caption{Overall architecture of \dacapo.}
    \vspace{-2ex}
    \label{fig:architecture-overview}
\end{figure}

Figure~\ref{fig:architecture-overview} presents the overall architecture of the \dacapo accelerator. 
The design is built upon a standard systolic array architecture, which is composed of two-dimensional processing engines.
There are two primary differences compared to conventional TPU-like systolic array architectures: (1) the accelerator is \emph{spatially} partitionable into Top Sub-Accelerator (T-SA) and Bottom Sub-Accelerator (B-SA); and (2) the processing engines are not simple multiplication-and-accumulation (MAC) units, but dot-product units that can perform vector dot products of 2-bit, 4-bit, and 8-bit operands, depending on the MX precision mode. 
This section will describe the two architectural techniques in more detail. 

\subsection{Spatially-Partitionable Architecture}
\label{subsec:spatially-partitionable-architecture}
\niparagraph{Why revisit spatially partitionable architectures?}
Spatial partitioning techniques for neural network accelerators have been explored in previous works, such as Planaria~\cite{planaria:2020} and Dataflow mirroring~\cite{dataflowmirroring:2021}.
Planaria utilizes architectural fission to dynamically compose processing elements into an arbitrary number of sub-accelerators, enabling efficient multi-DNN inference and adherence to service-level objectives (SLOs). 
However, Planaria's on-chip buffer partitioning and inter-buffer networks are quite complex.
Dataflow mirroring proposes a simpler four-way partitioning scheme for systolic arrays, prioritizing efficiency over flexibility.
While it provides simpler on-chip memory and networking architecture, it targets multi-tenant inference serving, which demands excessive adaptability than continuous learning.
To address this, we introduce a two-way spatial partitioning approach that effectively balances architectural simplicity with the necessary flexibility, efficiently facilitating the parallel execution of the three kernels.

\niparagraph{Row-granular processing element organization.}
For simplicity, we propose to group the two-dimensional processing elements of systolic arrays into one-dimensional \emph{rows}, which can be composed into two groups: T-SA and B-SA. 
In typical systolic arrays, the input, weight, and output tensors flow either horizontally or vertically, depending on which data is stationary. 
In designing the dataflow for \dacapo, leveraging the property that both weights and outputs flow vertically, we propose to enable vertical communication channels in ``both'' directions, placing weight and output buffers at both the top and bottom of the systolic array. 
This way, regardless of how the rows are assigned to T-SA and B-SA, weights and outputs can flow vertically, so that the two stacked SAs can simultaneously run independent matrix-matrix multiplications without interfering with the other section. 
While we can support both weight and output stationary designs, we employ output stationary design for our RTL prototype. 
\niparagraph{Programmable memory interface.}
The proposed spatial partitioning approach requires T-SA and B-SA to place tensors at the input (I) and weight (W) buffers differently, as their dataflow differs. 
We do not perform this tensor transformation in software, but a programmable memory interface orchestrates the memory layout. 
Once our resource allocation algorithm determines the row assignments for T-SA and B-SA, it also reprograms the memory interface to feed the operands into the on-chip buffers in a way that the sub-SAs can directly read. 
The memory controller is designed to support MX and tensor manipulation operations using MX metadata decoder and bitwise concatenator to produce DPE operands. 
Note that this programmable memory interface is largely based on the memory interface used by several prior work, including MX~\cite{mx-isca-2023}, FAST~\cite{fast:2022}, and BitFusion~\cite{bitfusion:isca18}. 
\subsection{Precision-Flexible Dot-Product Engine}
Another dimension of reconfiguration in the \dacapo accelerator is MX precision~\cite{mx-isca-2023}. 
While it is statically determined for each kernel, every processing element must support multiple MX precisions, since they can be assigned to different SAs.
We describe the microarchitectural design of these processing elements, which we call Dot-Product Engine (DPE).
\begin{figure}[t]
    \centering
    \includegraphics[width=0.85\linewidth]{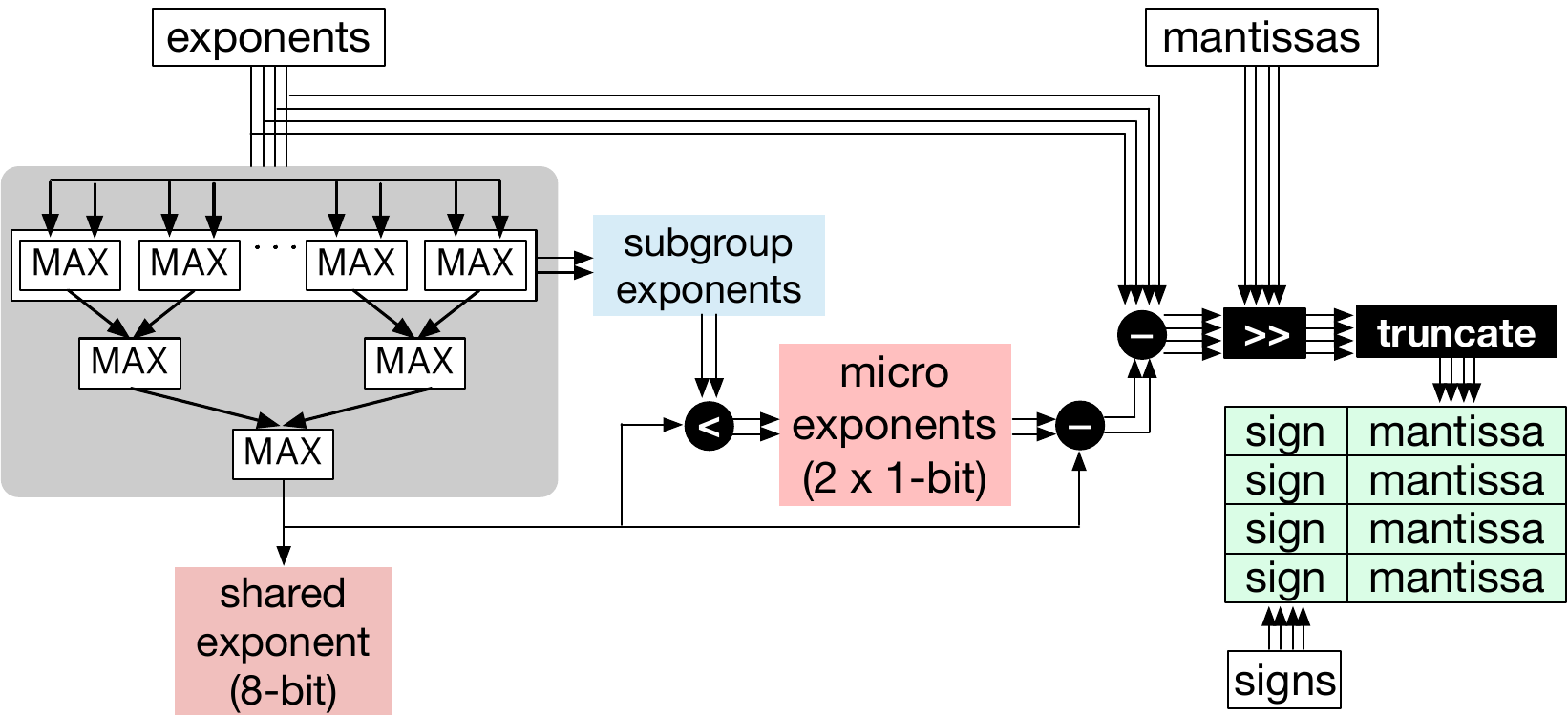}
    \caption{Conversion of four FP32 values into MX format.}
    \label{fig:MX_figure}
\end{figure}

\niparagraph{MicroeXponent (MX) format.} 
Figure~\ref{fig:MX_figure} illustrates how MX compresses a block of values differently from the conventional block floating point format.  
As in regular block floating point, MX compresses 16 address-adjacent values in memory as a block.
While the choice of block size is orthogonal to our DPE design, we use 16 as it is commonly used in most prior works~\cite{fast:2022,mx-isca-2023,flex-block:2022}.
However, for clearer illustration purposes, Figure~\ref{fig:MX_figure} employs a block size of 4.
MX first identifies the largest exponent within a block using a max tree and chooses it as the shared exponent.
Each block consists of sub-blocks where its values share a 1-bit microexponent (MX).
The sub-block size can vary, but we use 2, the same default value used in the original paper.
MX compares the maximum subgroup exponents obtained from the leaf stage of max tree with the shared exponent and sets the MX bit to 1 if every subgroup exponent is smaller than the shared exponent, effectively enlarging the supported dynamic range and thus enhancing accuracy resilience in lossy compression.
After determining the shared exponent and microexponents, mantissa values are shifted accordingly and at last, truncated from 23-bit of single-precision floating point to 2 (MX4), 4 (MX6), or 7 bits (MX9).

\begin{figure}[t]
    \centering
    \includegraphics[width=1.0\linewidth]{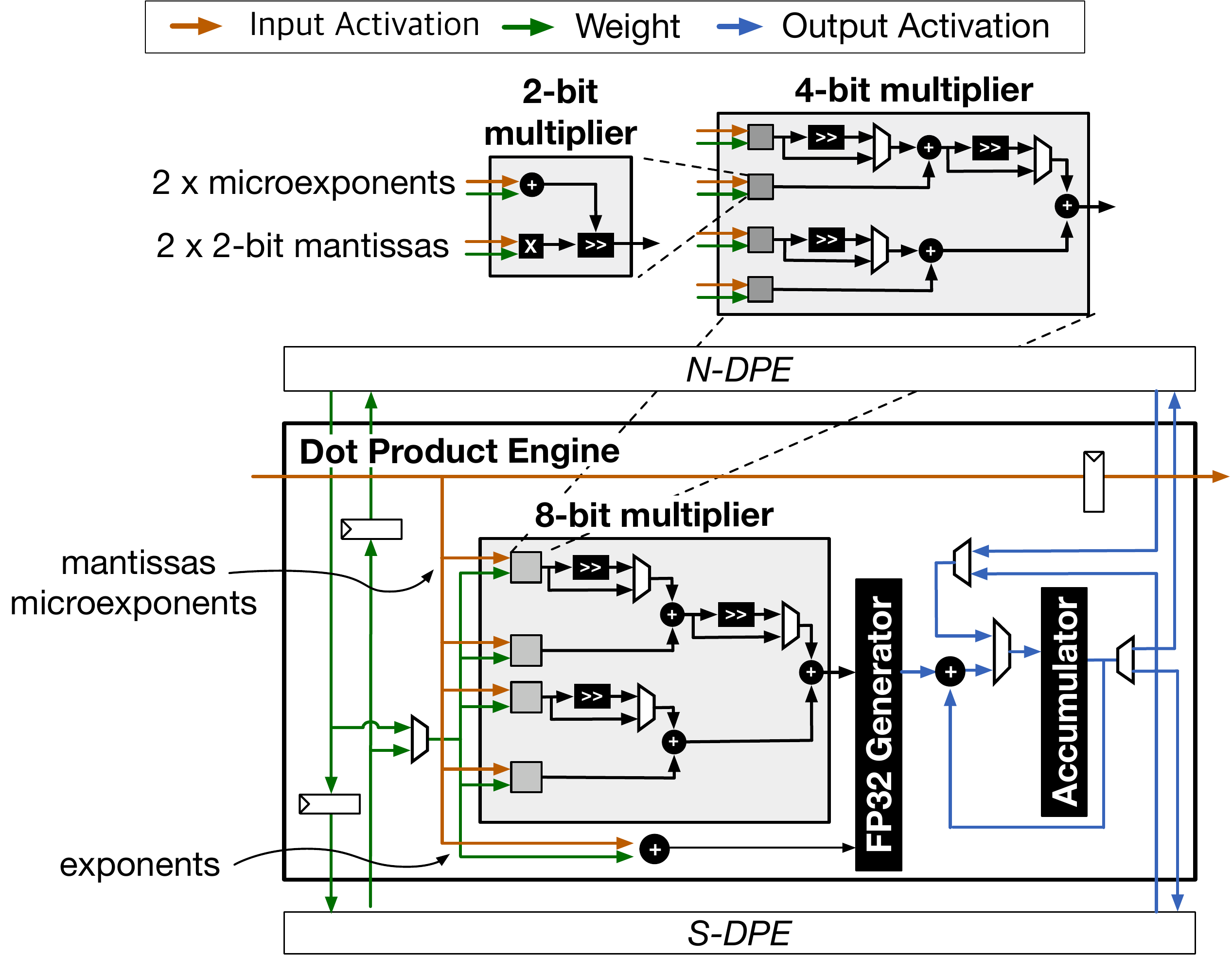}
    \caption{Microarchitecture of Dot-Product Engine (DPE).}
    \label{fig:dpe}
\end{figure}

\niparagraph{Microarchitectural opportunities from prior works.}
As the name implies, DPEs perform vector dot products between two MX-compressed vectors. 
The DPE microarchitecture is built upon techniques proposed in two prior works, Bit Fusion~\cite{bitfusion:isca18} and SIGMA~\cite{sigma:2020}. 
Bit Fusion~\cite{bitfusion:isca18} proposes bit-parallel dynamic fusion of small-bit width multiplications for quantized neural network inference.
SIGMA~\cite{sigma:2020}, on the other hand, proposes a forwarding adder network (FAN) that effectively reduces the results of multiple dot products using an adder tree structure equipped with a forwarding datapath. 
Inspired by these works, we leverage the proposed techniques in prior works for our DPE design, customizing them for performing dot products between MX-compressed vectors.  
\niparagraph{Dot-product engine: a hierarchical architecture for precision-flexible dot-product.}
Figure~\ref{fig:dpe} illustrates the microarchitectural details of a DPE.
To support various MX formats, we employ a hierarchical multiplication-and-accumulation (MAC) tree that can support 16$\times$2-bit (MX4), 4$\times$4-bit (MX6), or 1$\times$8-bit (MX9) in parallel. 
While the different MX modes require different numbers of mantissa values, the memory interface partitions the mantissa values in the granularity of 2-bits and supplies an array of 2-bits to the corresponding 2-bit multipliers. 
When it runs in the MX4 mode, every 2-bit multiplier performs independent multiplications, while for MX6 and MX9 mode, the results of 2-bit multipliers are accumulated by the hierarchical adder trees, producing 4-bit and 8-bit multiplication results, respectively. 

\niparagraph{Intra-DPE forwarding datapath and FP32 generator.}
Note that the multiplication results must be grouped in a specific size (e.g., 16) and then compressed into the MX format. 
Therefore, the accumulated results from the hierarchical MAC tree need to be collected at the FP32 generator, which requires direct result bypassing. 
To facilitate this, we devise a hierarchical result forwarding datapath, which can send multiplication outcomes through the MAC tree to the FP32 generator from any multiplier level.
The rate of FP32 result generation depends on the MX format.
For instance, MX4 completes all multiplications and accumulations for a dot product in just one cycle, whereas MX6 and MX9 require 4 and 16 cycles, respectively, due to the serial processing of subsets of the 16 multiplications (i.e., MX9: 16 cycles$\times$1 multiplication and MX6: 4 cycles$\times$4 multiplications). 
After the FP32 generator finishes the accumulations, it converts the final result into a floating-point value.
\niparagraph{Inter-DPE communication.}
Each DPE has bidirectional inter-DPE communication channels to its north and south.
It also has a horizontal channel, which pipelines the input activations from west to east, as in existing systolic array designs.
The two vertical channels are activated simultaneously.
One streams weight parameters to DPEs, while the other drains accumulated results to output buffers.
\subsection{Precision-Conversion Unit}
Note that when a set of results are produced by a SA and received by the output, they are floating-point values, not compressed in MX format. 
Therefore, the \dacapo accelerator comes with two precision-conversion units for the two SAs, which partition the floating-point output values into MX blocks of size 16. 
For inference and labeling, only row-major matrices are produced by the precision-conversion unit.
However, in the case of retraining, the precision-conversion unit carries out column-major MX conversion, in addition to the default row-major conversion, because transposed matrices are needed for gradient computation and weight updates. 
\section{Spatiotemporal Resource Allocation Algorithm}
\label{sec:resource}
This section describes \dacapo's resource allocation algorithm, which is pivotal for optimizing continuous learning.
\subsection{Algorithm Overview}
Our resource allocation algorithm operates on both spatial and temporal dimensions, determining spatial allocations statically during the offline phase and continuously adjusting temporal allocations dynamically in the online phase.
Unlike conventional fixed-window periodic retraining methods, \dacapo's temporal resource allocation mechanism promptly alternates between the retraining and labeling phases, with their durations governed by the algorithm-determined number of samples for each cycle.
At the end of each cycle, \dacapo assesses the impact of these phases on accuracy and adjusts temporal resource allocation dynamically.
This algorithm requires maintaining a fixed-capacity sample buffer to store labeled samples.
For retraining, a subset is drawn from this buffer, which is subsequently updated with newly labeled data in the following labeling phase.
At the end of retraining, \dacapo compares the validation accuracy of existing data with that of newly labeled data to detect data drift.
This facilitates \dacapo to make informed decisions regarding resource allocation for labeling in upcoming phases.
In the following section, we provide a formal description of the algorithm's operation.

\subsection{Hyperparameters for Resource Allocation}
\label{subsec:resource-variable}
\begin{table}[t]
\small
\caption{Notations of hyperparameters used in \dacapo's spatiotemporal resource allocation algorithm.}
\centering
\label{tab:variables}
\begin{tabular}{cl}
\hlinewd{1pt}
\multicolumn{2}{c}{\textbf{Retraining Hyperparameters}} \\
\hlinewd{1pt}
${N}_{t}$ & Number of samples for retraining \\
${N}_{v}$ & Number of samples for validation \\

\hlinewd{1pt}
\multicolumn{2}{c}{\textbf{Labeling Hyperparameters}} \\
\hlinewd{1pt}
${N}_{l}$ & Number of samples to label at usual \\
${N}_{ldd}$ & Number of samples to label at data drift \\

\hlinewd{1pt}
\multicolumn{2}{c}{\textbf{Spatial Allocation Hyperparameters}} \\
\hlinewd{1pt}
${R}_{tsa}$ & Number of DPE rows at T-SA \\
${R}_{bsa}$ & Number of DPE rows at B-SA \\

\hlinewd{1pt}
\end{tabular}
\end{table}
Table~\ref{tab:variables} shows the hyperparameters relevant to spatiotemporal resource allocation.
They are primarily concerned with the number of samples and influence the duration of retraining and labeling phases.
As \dacapo operates without a fixed window time, the values assigned to these hyperparameters dictate the scheduling of the retraining and labeling phases.

\niparagraph{Hyperparameters for retraining.}
For retraining, \dacapo loads a subset of samples from the sample buffer with buffer capacity ${C}_{b}$. 
The number of samples for retraining (${N}_{t}$) and validation (${N}_{v}$) are predefined, while ${N}_{v}$ is set to one third of ${N}_{t}$. 
These hyperparameters are decided according to the model size, as it has a direct impact on the computational cost required for retraining. 

\niparagraph{Hyperparameters for labeling.}
The quantity of samples for labeling is specified by two distinct hyperparameters: ${N}_{l}$ and ${N}_{ldd}$.
${N}_{l}$ specifies the standard number of samples to be labeled under normal conditions, while ${N}_{ldd}$ is an increased number of samples to be selected for labeling when data drift is detected.
In \dacapo, ${N}_{ldd}$ is established as a multiple of ${N}_{l}$. 
We empirically select this multiple to be four.

\niparagraph{Hyperparameters for spatial allocation.}
The number of DPE rows at T-SA and B-SA are defined as ${R}_{tsa}$ and ${R}_{bsa}$.
To optimize them, we prioritize the ${R}_{tsa}$ where the retraining and labeling phases run on, while ensuring that the ${R}_{bsa}$, the inference resources, is sufficient to meet the latency requirements of streaming input frames.
\SetKwInOut{Input}{input}
\SetKwInOut{Output}{output}
\SetKwComment{Comment}{\textsf{//} }{}
\SetKwComment{CommentWithSpace}{\quad\quad\textsf{//} }{}
\SetKwComment{CommentTwo}{/* }{*/}

\begin{algorithm}[tb]
\caption{Spatiotemporal Resource Allocation}
\label{alg:scheduling}

\KwIn{$
 \hspace{0.75em}{B}_{cur}: \textit{Current\enspace sample\enspace buffer} \newline 
 \hspace*{0.75em}{W}_{cur}: \textit{Current\enspace weight\enspace of\enspace student} \newline 
 \hspace*{0.75em}{C}_{b}: \textit{Capacity of sample buffer}\newline
 \hspace*{0.75em}{V}_{thr}: \textit{Threshold value to detect data drift}\newline
 \hspace*{0.75em}{N}_{t,v}: \textit{Retraining hyperparameters}\newline
 \hspace*{0.75em}{N}_{l,ldd}: \textit{Labeling hyperparameters}
$}

\SetKwFunction{GetSpatialAllocation}{\textnormal{GetSpatialAllocation}}
\SetKwFunction{SetAccelerator}{\textnormal{SetAccelerator}}
\SetKwFunction{GetData}{\textnormal{GetData}}
\SetKwFunction{Retrain}{\textnormal{Retrain}}
\SetKwFunction{UpdateWeight}{\textnormal{UpdateWeight}}
\SetKwFunction{Valid}{\textnormal{Valid}}
\SetKwFunction{Label}{\textnormal{Label}}
\SetKwFunction{GetInferOutputs}{\textnormal{GetInferOutputs}}
\SetKwFunction{Evaluate}{\textnormal{Evaluate}}
\SetKwFunction{ResetBuffer}{\textnormal{ResetBuffer}}
\SetKwFunction{UpdateBuffer}{\textnormal{UpdateBuffer}}
\vspace{1ex}

\Comment{\textit{\textsf{\small{Spatial resource allocation}}}}
${R}_{tsa}, {R}_{bsa} \gets \GetSpatialAllocation{}$\;
$\SetAccelerator{\ensuremath{{R}_{tsa}, {R}_{bsa}}}$\;
\vspace{1ex}

\Comment{\textit{\textsf{\small{Temporal resource allocation}}}}
\While {$true$}
{
    \Comment{\textit{\textsf{\small{Retraining}}}}
    ${D}_{t}, {D}_{v} \gets \GetData{\ensuremath{{B}_{cur}, {N}_{t}, {N}_{v}}}$\;
    ${W}_{nxt} \gets \Retrain{\ensuremath{{W}_{cur}, {D}_{t}}}$\; 
    ${W}_{cur} \gets \UpdateWeight{\ensuremath{{W}_{nxt}}}$\; 
    ${acc}_{v} \gets \Valid{\ensuremath{{W}_{cur}, {D}_{v}}}$\;
    \vspace{1ex}
    
    \Comment{\textit{\textsf{\small{Labeling}}}}
    ${D}_{l} \gets \Label{\ensuremath{{N}_{l}}}$\;
    ${O}_{prd} \gets \GetInferOutputs{\ensuremath{{D}_{l}}}$\;
    ${acc}_{l} \gets \Evaluate{\ensuremath{{O}_{prd}, {D}_{l}}}$\;
    \vspace{1ex}
    
    \Comment{\textit{\textsf{\small{Check data drift \& update Buffer}}}}
    \If{${acc}_{l} - {acc}_{v} < {V}_{thr}$}
    {
        $\ResetBuffer{\ensuremath{{B}_{cur}}}$\; 
        ${D}_{l} \gets \Label{\ensuremath{{N}_{ldd} - {N}_{l}}} + {D}_{l}$\;
    }
    
    ${B}_{cur} \gets \UpdateBuffer{\ensuremath{{B}_{cur}, {D}_{l}, {C}_{b}}}$\;
}
\end{algorithm}
\subsection{Spatiotemporal Resource Allocation Algorithm}
\label{subsec:resource-benefit}
Algorithm~\ref{alg:scheduling} describes the temporal resource allocation algorithm in detail.
\emph{Line 1-2}: \dacapo allocates resources for T-SA and B-SA during the offline phase.
\emph{Line 4-7}: In the retraining phase, \dacapo evaluates the updated model's weights (${W}_{nxt}$) using a validation dataset (${D}_{v}$) sampled from the current sample buffer (${B}_{cur}$).
Note that as validation only involves forward-pass processing on T-SA, its computational cost is significantly lower than that of retraining.
The validation accuracy assesses the new model's adaptability to the accumulated dataset.
\emph{Line 8-10}: During the labeling phase, \dacapo matches inference outputs with corresponding samples to evaluate the current model's inference accuracy on the newly sampled dataset.
\emph{Line 11}: If the validation accuracy subtracted from the labeling accuracy is lower than the threshold (${V}_{thr}$), it indicates data drift.
\emph{Line 12-13}: Upon detecting data drift, \dacapo clears the current sample buffer to remove outdated samples, and labeling time is extended to ensure that these better-suited samples are included in the dataset for subsequent phases.
\emph{Line 14}: Once all new samples are labeled, \dacapo updates its sample buffer.
With this algorithm, \dacapo adeptly detects and addresses data drift by dynamically allocating labeling time based on the validation accuracy from a segment of the sample buffer.
\subsection{Hyperparameter Tuning for Different Environments}
\label{subsec:variable-calibration-on-runtime}
As noted in Section~\ref{subsec:resource-variable}, we use a fixed set of empirically determined hyperparameters for the resource allocation algorithm. 
To determine the hyperparameters, we use the BDD100K dataset (see Section~\ref{subsec:methodology}) and exhaustively explore the hyperparameter search space to establish their values, consistently finding that the chosen settings outperform alternatives across various environmental scenarios, suggesting a robust insensitivity to scenario changes.
Given this empirical insight, we choose to use an offline hyperparameter tuning method, leaving the online hyperparameter auto-tuning to future work.  
Before deploying autonomous systems in real-world environments, \dacapo requires offline hyperparameter tuning to identify the optimal settings.
The tuning is required only once for each autonomous system, representing a cost that is not only inexpensive but also amortizable over time.
\section{Evaluation}
\label{sec:experiment}
\subsection{Methodology}
\label{subsec:methodology}
\begin{table}[t]
    \small
    \caption{Descriptions of workload scenarios for continuously learning video analytics on autonomous systems.}
    \label{tab:scenarios}
    \centering
    \begin{tabular}{ccc}
    \hlinewd{1.0pt}
    \textbf{Name} & \textbf{Weather} & \textbf{Data Drift Types (Only 1)} \\
    \hlinewd{1.0pt}
    S1 & Clear & \multirow{2}{*}{Label Distribution} \\
    S2 & Overcast &  \\
    \hline
    S3 & Clear & \multirow{2}{*}{Label Distribution, Time of Day} \\
    S4 & Snowy & \\
    \hline
    S5 & Clear & \multirow{2}{*}{Label Distribution, Time of Day, Location} \\
    S6 & Rainy & \\
    \hlinewd{1.0pt}
    \multicolumn{1}{c}{\textbf{Name}} & \multicolumn{2}{c}{\textbf{Composition of Data Drift Types (All 4)}} \\
    \hlinewd{1.0pt}
    \multicolumn{1}{c}{ES1} & \multicolumn{2}{c}{\multirow{2}{*}{Label Distribution, Time of Day, Location, Weather}}\\
    \multicolumn{1}{c}{ES2} &  \\
    \hlinewd{1.0pt}
    \end{tabular}
\end{table}

\begin{figure}[t]
    \centering
    \includegraphics[width=\linewidth]{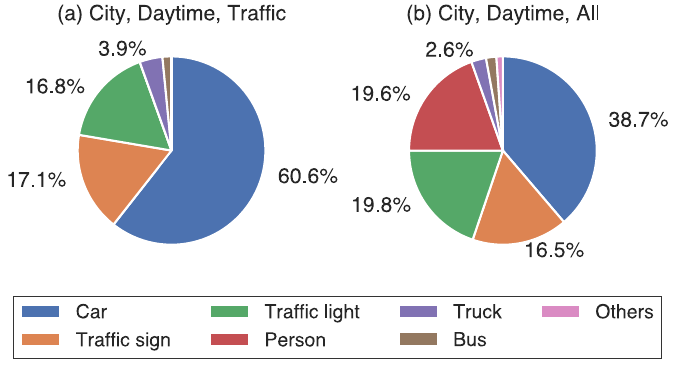}
    \caption{Different label distributions within distinct 60-second segments of an example scenario.}
    \label{fig:bdd100k-label-distribution}
\end{figure}

\niparagraph{Dataset and data drifts.}
We use BDD100K~\cite{bdd100k}, a comprehensive live video driving dataset with an abundance of varied situations.
To adapt it for classification, we crop individual objects within the video frames and arrange them in chronological order.
To create data drifts, we characterize videos based on three key attributes: Label Distribution, Time of Day, and Location.
There are two types of ``Label Distribution'': (1) \emph{Traffic Only} which focuses on traffic-related labels, and (2) \emph{All} which includes a broader range of labels, such as pedestrian, bicycle, and motorcycle.
``Time of Day'' is classified into \emph{Daytime} and \emph{Night} to account for varying lighting conditions.
``Location'' is divided into \emph{City} and \emph{Highway}, each presenting unique driving environments.
This setup facilitates the evaluation of \dacapo's performance in various data drift situations, providing insights into its flexibility and adaptability.

\niparagraph{Scenarios.}
Every scenario constructed for evaluation is created by concatenating a series of video clips.
Each video is treated as a collection of frames within the scenario, unfolding over a duration of 20 minutes, at a frame rate of 30 FPS.
We develop three distinct types to capture diverse data drift scenarios including (1) 6 regular scenarios with single data drift at a time (S1--S6), and (2) 2 extreme scenarios with 4 data drifts occurring at the same time (ES1 and ES2). 
Table~\ref{tab:scenarios} provides the details of 8 scenarios and Figure~\ref{fig:bdd100k-label-distribution} shows the label distributions present across different segments of scenarios. 
\begin{table}[t]
    \small
    \caption{Specifications of the evaluated DNN models.}
    \label{tab:benchmark-model}
    \centering
    \begin{tabular}{cccc}
    \hlinewd{1.0pt}
    \textbf{Type} & \textbf{Name} & \textbf{Parameters} & \textbf{GFLOPs} \\
    \hlinewd{1.0pt}
    \multirow{3}{*}{Student}
    & ResNet18 & 11.7M & 1.82 \\
    & ResNet34 & 21.8M & 3.67 \\
    & ViT-B/32 & 88.2M & 4.37 \\
    \hlinewd{0.5pt}
    \multirow{3}{*}{Teacher}
    & WideResNet50 & 68.9M & 11.43 \\
    & ViT-B/16 & 86.6M & 16.87 \\
    & WideResNet101 & 126.9M & 22.80 \\
    \hlinewd{1.0pt}
    \end{tabular}
\end{table}
\niparagraph{DNN models.}
We evaluate \dacapo's performance on object classification tasks using six DNN models listed in Table~\ref{tab:benchmark-model}.
These models are paired as student and teacher models: (1) ResNet18 and WideResNet50, (2) ViT-B/32 and ViT-B/16, (3) ResNet34 and WideResNet101.

\niparagraph{Accuracy metric.}
\dacapo differs from other baselines as it does not utilize a window time.
For a fair comparison, we assess its averaged accuracy in the scenarios over time slices corresponding to the window period.

\niparagraph{Retraining and inference hyperparameters.}
The learning rate of retraining is set to $10^{-3}$, and we use SGD optimization with batch size of 16.
For inference, we use batch size 1.
While the \dacapo hardware supports MX4, MX6, and MX9 precision modes, we use MX9 (7-bit mantissas) for retraining and MX6 (4-bit mantissas) for inference.
\niparagraph{Hardware development and synthesis.}
Our \dacapo prototype constitutes of a systolic array with 16$\times$16 DPEs and a 96KB on-chip SRAM. 
Note that while our prototype is designed as a tiny chip, \dacapo could scale the number of DPEs to larger configurations (e.g., 32$\times$32) or multiple \dacapo chiplets could be packaged together if there is a need. 
For evaluation, we implement Verilog RTL design, and verify the results through an RTL simulator.
We synthesize in 28nm CMOS technology using Synopsys Design Compiler and CACTI, which provide the dynamic/static power and area of each component of \dacapo, as reported in Table~\ref{tab:hardware-area-power}.
We also develop an in-house cycle-accurate software simulator to measure the execution cycles of DPE cores.
\niparagraph{GPU.} We evaluate \dacapo against NVIDIA Jetson Orin, a GPU commonly used in autonomous system computing environments, specifications of which are provided in Table~\ref{tab:hardware-area-power}.
Out of the multiple power options provided, we use two options: high power (default power settings) and low power (30W power constraints) in our evaluations, to understand the behavior of our workload on low-power GPUs.

\niparagraph{System simulator implementation.}
To measure end-to-end accuracy in continuous learning with hardware-based retraining and labeling phases, we develop a system simulator, which uses the accelerator simulator results.
It simulates the runtime behaviors of a \dacapo-equipped autonomous system, while the accuracy evaluation is performed on the GPUs. 
Prior to starting continuous learning, the system generates DNN cycle statistics for the partitioned systolic array, informing the system about the samples processed per phase within a given timeframe. 
This information guides the execution of continuous learning jobs on the GPU, aligning with the scheduled hardware allocation.
By integrating hardware simulation and GPU kernel execution, our approach simulates end-to-end accuracy and the spatiotemporal allocation algorithm within our continuous learning system.
\begin{table}[t]
    \small
    \caption{Evaluated GPU and \dacapo platforms.}
    \label{tab:hardware-area-power}
    \centering
    \begin{tabular}{c|c|c}
    \hlinewd{1.0pt}
    \textbf{Device} & \textbf{Jetson Orin} & \dacapo\\
    \hlinewd{1.0pt}
    Technology & 8 nm & 28 nm  \\
    \hlinewd{0.5pt}
    Area & N/A & 2.501 mm² \\
    \hlinewd{0.5pt}
    Frequency & 1.3 GHz & 500 MHz\\
    \hlinewd{0.5pt}
    Power & 15 - 60 W &  0.236 W \\
    \hlinewd{0.5pt}
    \makecell{DRAM \\ Bandwidth} & \makecell{LPDDR5 \\ 204.8 GB/s} & \makecell{LPDDR5 \\ 204.8 GB/s} \\
    \hlinewd{1.0pt}
    \end{tabular}
\end{table}

\begin{figure*}[t]
    \centering
    \includegraphics[width=\linewidth]{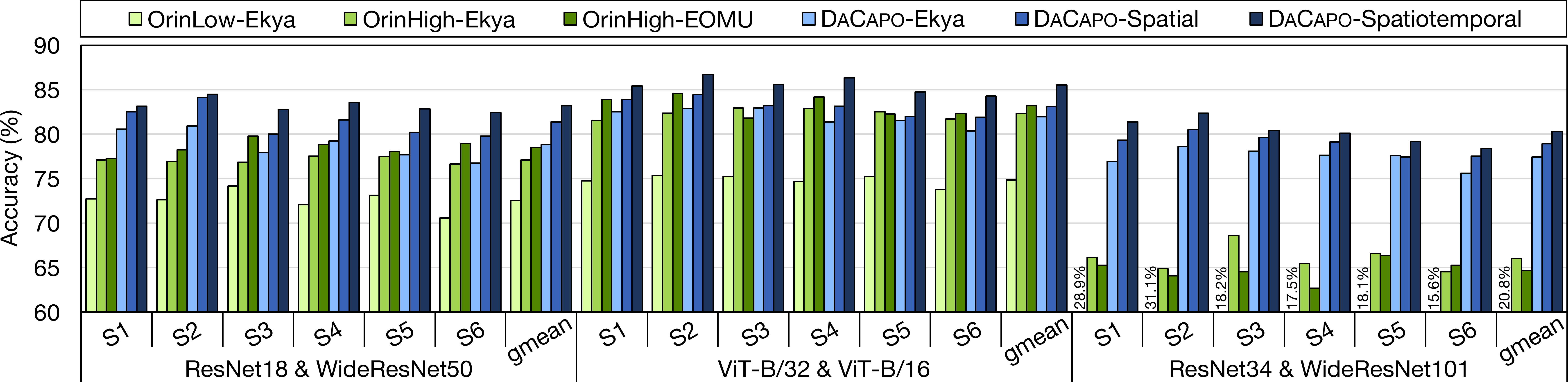}
    \caption{End-to-end averaged accuracy of different continuously learning autonomous systems on six runtime scenarios.}
    \label{fig:eval-accuracy}
\end{figure*}

\niparagraph{Baselines.} 
We compare two \dacapo systems with several baseline variants, represented by the combinations of their hardware platforms and resource allocation methods. 
For both baseline and \dacapo variants, we use two hardware platforms, (1) NVIDIA Orin GPU, and (2) our \dacapo accelerator. 
We select two state-of-the-art continuous learning systems as the baseline: (1) Ekya~\cite{ekya-nsdi-2022} sets the retraining configuration through a profiling process at every window, and (2) EOMU~\cite{eomu-mm-2023} selectively triggers the retraining within a shorter window and determines the training configurations by the retraining manager. 
Below, we describe all the compared continuous learning systems categorized by their combinations of hardware and resource allocation techniques. 
\begin{itemize}
\item \textbf{OrinLow-Ekya.} 
With this baseline, we examine how power levels influence Ekya's performance on autonomous systems in low battery conditions.
Orin supports multiple power options, ranging from 15W to 60W.
``OrinLow'' represents Orin using the 30W option, which consumes 127$\times$ more power than \dacapo.
This baseline sets the GPU's maximum frequency to 624.8MHz, the closest to \dacapo's 500MHz.
\item \textbf{OrinHigh-Ekya.} 
We evaluate the performance of Ekya on Orin's default power setting (i.e., 60W) without power constraints, comparing it to \dacapo on a sufficiently powered autonomous system.
This baseline consumes 254$\times$ more power than \dacapo.
\item \textbf{OrinHigh-EOMU.} 
We assess the performance of EOMU on Orin GPU at the default power setting. 
This baseline uses shorter window times to detect data drift and trigger the retraining process. 
We set 10 seconds as the window time for EOMU according to its paper. 
\item \textbf{\dacapo-Ekya.}
This baseline represents a \dacapo-accelerated autonomous system using Ekya's resource allocation and system configuration mechanism. 
\end{itemize}
\subsection{Accuracy Comparison}
\label{subsec:evaluation_accuracy-comparison}

Figure~\ref{fig:eval-accuracy} presents a comparison of the end-to-end accuracy results among the aforementioned continuous learning system variants. 
We compare the baselines to the two \dacapo variations with different resource allocation methods: (1) \dacapo-Spatial: optimized static resource allocation within a predetermined window time, and (2) \dacapo-Spatiotemporal: dynamic spatiotemporal allocation responsive to data drift. 

\niparagraph{OrinLow-Ekya.}
When we use Orin with the low-power mode, the entirety of three pairs we used for evaluation offer significantly low accuracy, which ranges from 20.8\% to 74.8\%. 
For all experiments, we prioritize resource allocation for inference, incurring no frame drops that largely affect the final accuracy. 
Therefore, the limited accuracy results is solely attributed to insufficient computation resources offered for retraining and labeling. 
In particular, the (ResNet34, WideResNet101) pair demonstrates notably lower accuracy compared to the others since the relatively large model size requires a substantial portion of the Orin GPU's computational resources for inference, leaving minimal resources available for the other two kernels.
\niparagraph{OrinHigh-Ekya and OrinHigh-EOMU.}
We also compare the Orin GPU baselines with the highest power setting, OrinHigh-Ekya and OrinHigh-EOMU, to examine the full capabilities of NVIDIA's autonomous system solution. 
While the two obtain a significant accuracy improvement of respective 11.5\% and 13.0\% in comparison with OrinLow-Ekya, it is still not closely reaching \dacapo-Spatiotemporal, which suggests potential unleashed benefits of the proposed acceleration solution. 
Intuitively, when evaluated on the toughest case of (ResNet34, WideResNet101) pair, the two see the largest gains of 45.2\% and 43.9\% improvements, respectively. 
We observe a slight difference between Ekya and EMOU since both are based on an assumption that labeling and training are offloaded to remote servers, which makes EOMU's improved resource allocation method marginally influential on the end accuracy. 
\niparagraph{\dacapo-accelerated systems.}
To better understand the implication of the resource allocation algorithm on the \dacapo hardware, we populate three \dacapo-based system variants using our Spatial and Spatiotemporal resource allocation methods along with Ekya. 
While the \dacapo accelerator offers significantly higher raw performance, we observe that \dacapo-Ekya does not always outperform the Orin baselines significantly. 
This is because Orin uses significantly higher precision (i.e., 32-bit singe-precision FP), while \dacapo leverages low precision (i.e., MX4 to MX9) for higher throughput, which can negatively impact the accuracy behaviors depending on the used models. 
Particularly for the ViT models, we notice that \dacapo-Ekya performs poorly even compared to the Orin baselines since ViT models often have relatively larger precision sensitivity than other models~\cite{ptq-for-vit-neurips-2021, ptq4vit-eccv-2022}.
Similarly, \dacapo-Spatial offers superior performance compared to the Orin baselines, while the gap is marginal. 
On the other hand, when employing \dacapo-Spatiotemporal, the accuracy progresses towards the highest level among all other continuous learning systems, akin to completing the ``last mile'' in a journey towards optimal performance under the computational resources of the \dacapo hardware. 
Concretely, \dacapo-Spatiotemporal achieves significant accuracy gains over OrinHigh-Ekya and OrinHigh-EOMU, which are respectively 6.5\% and 5.5\% improvements. 
The results suggest that our hardware-algorithm co-designed acceleration approach effectively unlocks continuous learning's full capabilities and successfully addresses the ``resource wall'' challenge. 

\begin{figure*}[t]
    \centering
    \includegraphics[width=0.9\linewidth]{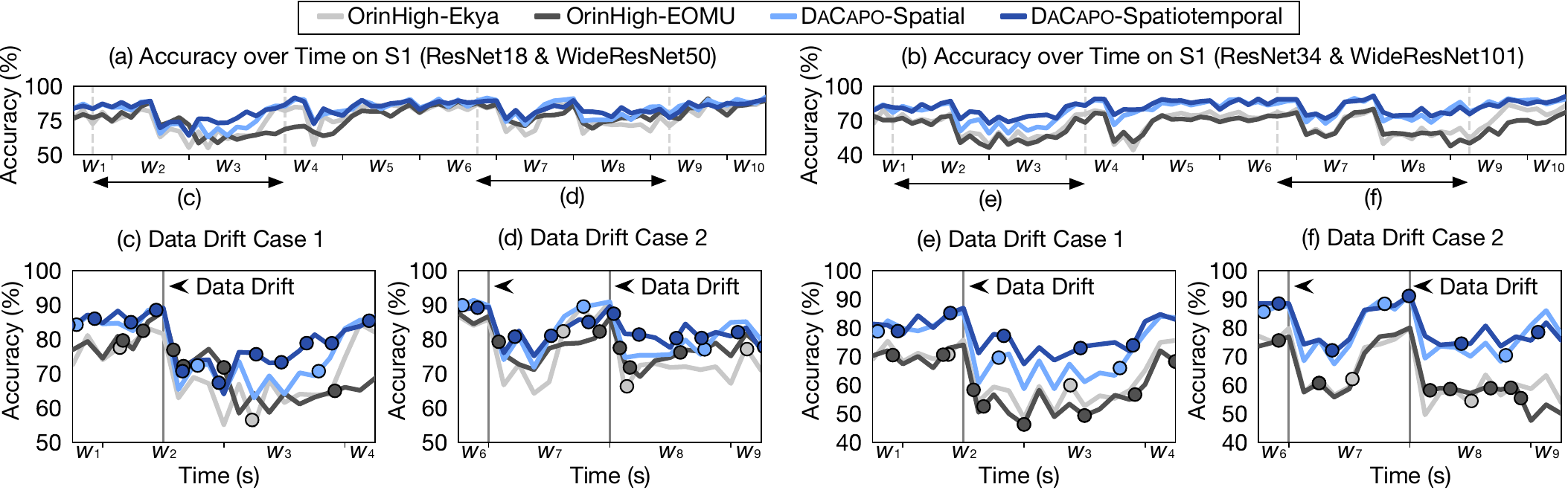}
    \caption{Accuracy over time. To illustrate the time-series accuracy pattern, we display the average accuracy recorded at 15-second intervals. Each colored dot on the line indicates the completion of a retraining phase.}
    \label{fig:over-time}
\end{figure*}

\subsection{In-Depth Analysis on Resource Allocation Algorithm}
\label{subsec:evaluation_systemwide-accuracy-over-time}
We now delve into the in-depth analysis of accuracy change over time in a specific experiment using scenario S1. 

\niparagraph{Accuracy over time.}
To better analyze the effectiveness of \dacapo-Spatiotemporal over \dacapo-Spatial and the Orin baselines, Figure~\ref{fig:over-time}(a) and Figure~\ref{fig:over-time}(b) show accuracy change over an extended timeframe, using (ResNet18, WideResNet50) and (ResNet34, WideResNet101) model pairs.
In both pairs, \dacapo-Spatiotemporal works better than \dacapo-Spatial and the Orin baselines as they lack ability to promptly adapt to data drift occurrences. 
The Ekya and \dacapo-Spatial systems use a fixed predetermined window time, which lacks situational awareness within the dataset. 
However, \dacapo-Spatiotemporal shows better adaptability to environment changes through iterative retraining as required.
EOMU is designed to trigger frequent retrainings, showing comparatively better accuracy behaviors than Ekya, while the Orin's resource insufficiency results in its limited gain. 
\niparagraph{Data drift cases.}
To analyze the accuracy changes in finer detail, Figure~\ref{fig:over-time}(c, d) and Figure~\ref{fig:over-time}(e, f) zoom in on specific intervals of interest in Figure~\ref{fig:over-time}(a) and Figure~\ref{fig:over-time}(b), respectively, allowing for a more focused analysis of accuracy variations observed in these time frames. 
Figure~\ref{fig:over-time}(c) and Figure~\ref{fig:over-time}(e) illustrate instances of data drifts where \dacapo-Spatiotemporal demonstrates its effectiveness, swiftly recovering from these drifts via regular retraining sessions.
This results in superior performance compared to \dacapo-Spatial, with maximum improvement margins of 13.1\% and 12.4\% for the two pairs, respectively.
Conversely, Figure~\ref{fig:over-time}(d) and Figure~\ref{fig:over-time}(f) illustrate two suboptimal cases where \dacapo-Spatial outperforms \dacapo-Spatiotemporal.
During certain periods, \dacapo-Spatial exhibits higher accuracy, with maximum margins of 5.5\% and 5.3\%, respectively.
However, the overall accuracies of \dacapo-Spatiotemporal and \dacapo-Spatial from $w_7$ to $w_8$ stand at 82.4\% and 81.7\% in Figure~\ref{fig:over-time}(d), and 79.7\% and 78.0\% in Figure~\ref{fig:over-time}(f), respectively.
Even in its suboptimal cases, \dacapo-Spatiotemporal maintains competitive performance with \dacapo-Spatial.
Additionally, Figure~\ref{fig:over-time}(a) and Figure~\ref{fig:over-time}(b) also delineate that the Orin baselines almost consistently provide lower accuracy than the \dacapo variants. 
In fact, the result shows that EOMU triggers substantially more frequent retrainings, as marked with the dark grey colored dots. 
However, such frequent retrainings do not always help because trainings with insufficent resource engender incomplete models, which ends up lowering the  final accuracy. 
Overall, the results suggest that \dacapo-Spatiotemporal outperforms all the alternative baselines by leveraging significantly boosted performance of hardware-accelerated computing platform as well as judiciously allocating resources in the spatiotemporal dimension. 
\subsection{Effectiveness of Temporal Resource Allocations}
\begin{figure}[t]
    \centering
    \includegraphics[width=\linewidth]{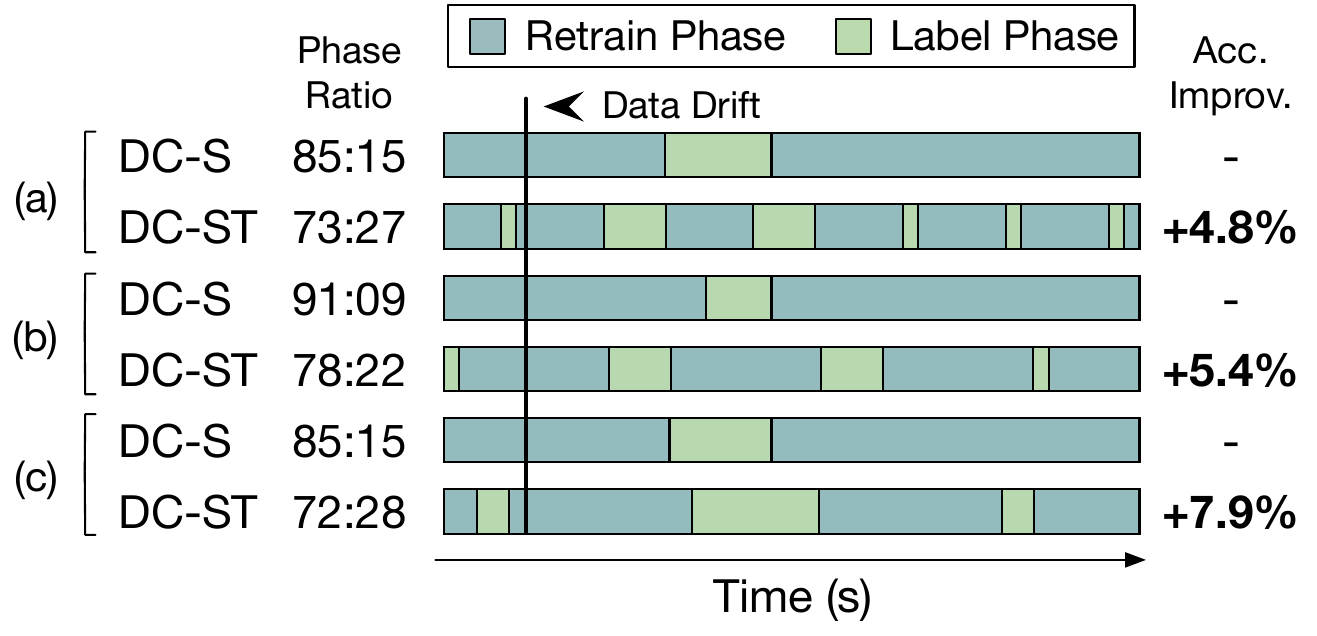}
    \caption{Scheduling decisions of temporal resource allocation over time: (a) ResNet18 \& WideResNet50, (b) ViT-B/32 \& ViT-B/16, and (c) ResNet34 \& WideResNet101. DC-S is \dacapo-Spatial and DC-ST is \dacapo-Spatiotemporal.}
    \label{fig:scheduling-decision}
\end{figure}
Figure~\ref{fig:scheduling-decision} shows the results of proposed temporal resource allocations, which are collected during 3 minutes on S1 using the evaluated three model pairs.
For each pair, we present the time breakdown of retraining and labeling phases for \dacapo-Spatial (DC-S) and \dacapo-Spatiotemporal (DC-ST).
Additionally, we report the overall breakdown of retraining and labeling on the left and the accuracy improvement results on the right. 
We observe that when data drifts occur, \dacapo-Spatiotemporal swiftly detects them and allocates 12.7\% higher time slots for labeling, achieving 5.9\% higher accuracy improvement compared to the spatial-only baseline. 
These results demonstrate the effectiveness of our strategy that allocates more resources to labeling at the occurrence of data drifts. 
Note that we achieve the improved accuracy even with ``less'' retraining time since we cut the assigned time for retraining, exhibiting the effectiveness of rebalancing. 
\subsection{Sensitivity Study for Extreme Data Drift Scenarios}
\label{subsec:extreme-scenario}
\begin{figure}[t]
    \centering
    \includegraphics[width=\linewidth]{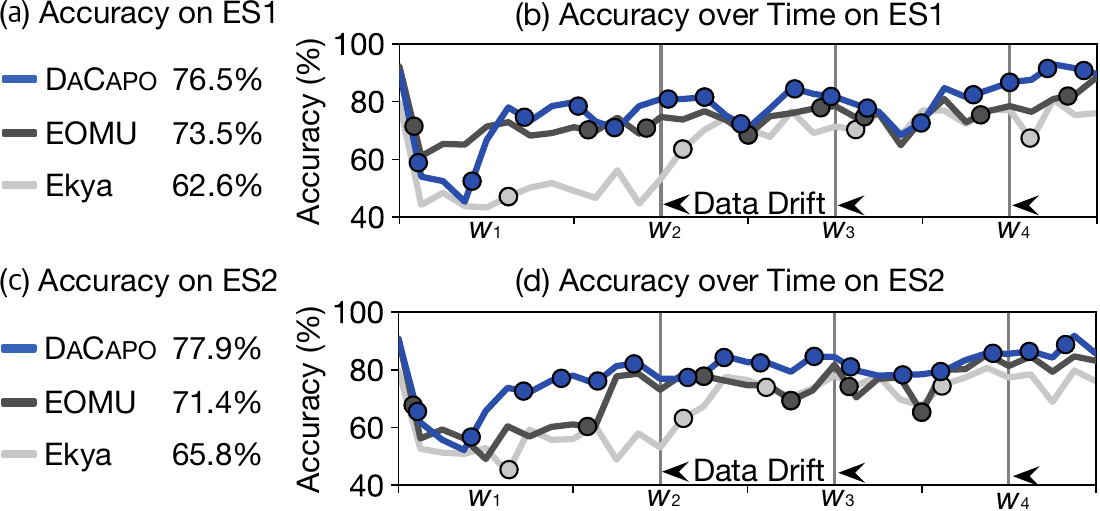}
    \caption{Accuracy comparison of \dacapo, EOMU, and Ekya, using two extreme data drift scenarios. We use (ResNet18 \& WideResNet50) model pair for this experiment.}
    \label{fig:extreme-scenario}
\end{figure}
Figure~\ref{fig:extreme-scenario} compares the accuracy results of three continuous learning systems, when we employ two extreme data drifts scenarios (ES1 and ES2) and a model pair (ResNet18 \& WideResNet50). 
In addition to the two extreme scenarios, we evaluate four more extreme scenarios and observe the similar trends, while we only show two cases due to the space limit. 
The figure reports the averaged accuracies (left) along with the accuracy change over time (right). 
Compared to the regular scenario results, Ekya experiences an accuracy degradation of 12.9\%.
Conversely, EOMU exhibits a higher tolerance than Ekya and offers 7.8\% higher accuracy. 
This improvement is attributed to EOMU's capability to monitor accuracy changes across short intervals and trigger adaptations by retraining models more frequently than Ekya, as illustrated by dense markers on the line graphs. 
However, \dacapo, by scheduling spatiotemporal resources for retraining and labeling, achieves the accuracy of 77.2\%, which is 4.4\% and 13.0\% higher than EOMU and Ekya, respectively. 
\section{Related Work}
\label{sec:related}

\niparagraph{Continuous learning.}
Continuous learning addresses the critical issue of deployed models becoming outdated over time due to data drift.
In the field of machine learning, significant efforts have been directed toward retraining pretrained models with new data without forgetting prior knowledge.
The replay-based approach~\cite{shim-cl-21-aaai, yoon-cl-22-iclr, icarl-cl-17-cvpr, rainbow-cl-21-cvpr, coresets-cl-20-nips} is one such method, integrating valuable historical datasets with new samples for model updates.
Alternatively, regularization-based updates~\cite{lee-cl-reg-20-cvpr, afec-cl-reg-21-nips, jung-cl-reg-20-nips, paik-cl-reg-20-aaai, sarwar-cl-reg-19-paml, kirkpatrick-cl-reg-17-pnas} augment additional regularization terms during further training to penalize updates on weights critical for previous tasks.
However, implementing these solutions often presents practical challenges due to unaddressed training overheads~\cite{prabhu-cl-23-cvpr, ghunaim-cl-23-cvpr}.
Distinctively, \dacapo adopts a strategy commonly employed in data streaming scenarios, as exemplified by the video analytics domain.
\niparagraph{Live video analytics.}
Edge devices, positioned near data sources, offer significant benefits, such as bandwidth conservation, reduced latency, and improved privacy by eliminating the need for data transmission to the cloud.
However, due to their limited computational capacity, most contemporary edge video analytics systems rely heavily on cloud infrastructure.
Prior works~\cite{hpca-2024-usas, dac-2023-shoggoth, dao-cl-filter-17-mass, wang-cl-filter-18-sec, ffsva-cl-filter-18-icpp, guo-eva-crop-19-tomm, chen-eva-crop-16-bigmm} aim to use edge devices for preprocessing tasks, such as filtering irrelevant frames or cropping the regions of interest.
Several studies~\cite{atc-2023-oakestra, www-2023-edgemove, iot-jia-2022, li-eva-part-16-sec, hung-eva-part-18-sec, lavea-eva-part-17-sec, liu-eva-part-18-edgesys, neurosurgeon-eva-part-17-asplos} propose partitioning DNN computations between edge devices and the cloud.
Other works~\cite{active-cl-sensys-2022, drolia-eva-reuse-17-sec, chen-eva-reuse-15-sensys, guo-eva-reuse-18-mobicom, drolia-eva-reuse-17-icdcs} propose enhancing efficiency by caching and reusing previous inference results.
Despite these attempts, the involvement of cloud computing constrains the full potential benefits of edge computing, explaining the increasing development of edge-targeted accelerators in the industry~\cite{tpu:edge, nvidia-jetson, tesla-20-micro, hailo-8, arm-ethos, xilinx-versal}.
Furthermore, while there are prior works that propose edge-targeted training solutions~\cite{hpca-2024-camel, carm-cl-22-dac, mobicom-2023-miro, cvpr-2023-mobilevos, aaai-2023-mobiletl, emsoft-2023-dacapo, cvpr-2022-rep, neurips-2020-tinytl}, they do not take into account the the tradeoff between resource utilization and final accuracy of training and labeling computations. 
\dacapo addresses these issues by a tailored hardware accelerator design that enables autonomous systems to perform independent retraining, thus facilitating a novel continuous learning video analytics system.
\niparagraph{Low-precision arithmetic for DNN acceleration.}
Replacing resource-intensive floating point operations with low-precision counterparts is a crucial optimization strategy for DNNs.
Notably, floating point representations with reduced bit width, such as FP16~\cite{micikevicius:mixed-precision:iclr:2018, fp16-17-hpec}, FP8~\cite{fp8-18-nips, fp8-19-nips}, BF16~\cite{bfloat16-google:2019}, and TF32~\cite{nvidia-tf32}, introduce certain improvements.
However, the inherent computational complexity of floating point often limits their effectiveness.
%
Conversely, fixed-point quantization techniques~\cite{dorefa, lin-quant-16-icml, tenary-quant-17-iclr, binarized-quant-16-nips, courbariaux-quant-15-nips, binary-tenary-quant-21-cvpr} present computational advantages but often lead to substantial degradation in model accuracy.
Block Floating Point (BFP) serves as an effective compromise between these extremes.
Courbariaux et al.~\cite{courbariaux-iclr-15} and Flexpoint~\cite{flexpoint:2017} introduce approaches to adjust shared exponent values either at fixed intervals or throughout the training process.
HybridBFP~\cite{hbfp:2018} calculates a shared exponent value for each dot-product operation, enhancing compressibility and accuracy.
MSFP~\cite{ms-fp:darvish:nips:2020} underscores the significant role of group sizes and validates their approach on production-level models.
FAST~\cite{fast:2022} utilizes stochastic rounding and a bit-serial hardware architecture to support dynamic bit width adjustment during training.
MX~\cite{mx-isca-2023} introduces additional micro-exponent bits to the shared exponent, showcasing enhanced performance across various domains.
\dacapo employs the microexponent concept from MX, effectively fulfilling the need for dynamic precision inference in our novel continuous learning system. 
\section{Conclusion}
\label{sec:conclusion}
Most existing continuous learning systems focus solely on training, neglecting the critical yet compute-heavy tasks of inference and labeling, and typically assume the availability of substantial computation resources, which is often impractical on battery-operated, resource-constrained autonomous systems.
To tackle this challenge, this work proposes a hardware-algorithm co-designed solution, \dacapo, which enables concurrent executions of retraining, labeling, and inference in a resource-efficient manner. 
\dacapo achieves this goal by jointly leveraging a spatially-partitionable, precision-flexible accelerator and a spatiotemporal resource allocation algorithm. 
Our empirical analyses across six real-world scenarios with varying data drifts demonstrate that \dacapo obtains accuracy improvements of 6.5\% and 5.5\% over the state-of-the-art systems, Ekya and EOMU, respectively, while consuming 254$\times$ less energy.
These results confirm that \dacapo is an effective and vital initial step to realize continuous learning on autonomous systems.
\section*{Acknowledgments}
We would like to thank Cliff Young,  Berkin Akin, and our anonymous reviewers for their review and valuable feedback on the paper.
This work was supported by Institute of Information \& communications Technology Planning \& Evaluation (IITP) (No.2024-00396013, No.2022-0-01037, No.2018-0-00503) under the Graduate School of Artificial Intelligence Semiconductor (IITP-2024-RS-2023-00256472), Information Technology Research Center (ITRC) support program (IITP-2024-2020-0-01795), and Artificial Intelligence Graduate School Program (KAIST) (No.2019-0-00075), funded by the Korea government (MSIT). 
%


\bibliographystyle{IEEEtranS}
\bibliography{paper}

\begin{thebibliography}{100}
\providecommand{\url}[1]{#1}
\csname url@samestyle\endcsname
\providecommand{\newblock}{\relax}
\providecommand{\bibinfo}[2]{#2}
\providecommand{\BIBentrySTDinterwordspacing}{\spaceskip=0pt\relax}
\providecommand{\BIBentryALTinterwordstretchfactor}{4}
\providecommand{\BIBentryALTinterwordspacing}{\spaceskip=\fontdimen2\font plus
\BIBentryALTinterwordstretchfactor\fontdimen3\font minus
  \fontdimen4\font\relax}
\providecommand{\BIBforeignlanguage}[2]{{%
\expandafter\ifx\csname l@#1\endcsname\relax
\typeout{** WARNING: IEEEtranS.bst: No hyphenation pattern has been}%
\typeout{** loaded for the language `#1'. Using the pattern for}%
\typeout{** the default language instead.}%
\else
\language=\csname l@#1\endcsname
\fi
#2}}
\providecommand{\BIBdecl}{\relax}
\BIBdecl

\bibitem{xilinx-versal}
S.~Ahmad, S.~Subramanian, V.~Boppana, S.~Lakka, F.-H. Ho, T.~Knopp, J.~Noguera,
  G.~Singh, and R.~Wittig, ``Xilinx first 7nm device: Versal ai core
  (vc1902).'' in \emph{HotChips}, 2019.

\bibitem{rainbow-cl-21-cvpr}
J.~Bang, H.~Kim, Y.~Yoo, J.-W. Ha, and J.~Choi, ``Rainbow memory: Continual
  learning with a memory of diverse samples,'' in \emph{CVPR}, 2021.

\bibitem{atc-2023-oakestra}
G.~Bartolomeo, M.~Yosofie, S.~B{\"a}urle, O.~Haluszczynski, N.~Mohan, and
  J.~Ott, ``Oakestra: A lightweight hierarchical orchestration framework for
  edge computing,'' in \emph{USENIX ATC}, 2023.

\bibitem{neuos}
S.~Bateni and C.~Liu, ``{NeuOS}: A {Latency-Predictable} {Multi-Dimensional}
  optimization framework for {DNN-driven} autonomous systems,'' in \emph{USENIX
  ATC}, 2020.

\bibitem{ekya-nsdi-2022}
R.~Bhardwaj, Z.~Xia, G.~Ananthanarayanan, J.~Jiang, Y.~Shu, N.~Karianakis,
  K.~Hsieh, P.~Bahl, and I.~Stoica, ``{Ekya: Continuous Learning of Video
  Analytics Models on Edge Compute Servers},'' in \emph{NSDI}, 2022.

\bibitem{coresets-cl-20-nips}
Z.~Borsos, M.~Mutny, and A.~Krause, ``Coresets via bilevel optimization for
  continual learning and streaming,'' \emph{NIPS}, 2020.

\bibitem{AU-Air}
I.~Bozcan and E.~Kayacan, ``{AU-AIR: A Multi-modal Unmanned Aerial Vehicle
  Dataset for Low Altitude Traffic Surveillance},'' in \emph{ICRA}, 2020.

\bibitem{9562043}
I.~Bozcan and E.~Kayacan, ``Context-dependent anomaly detection for low
  altitude traffic surveillance,'' in \emph{ICRA}, 2021.

\bibitem{ral-2021-gridnet}
I.~Bozcan, J.~Le~Fevre, H.~X. Pham, and E.~Kayacan, ``Gridnet: Image-agnostic
  conditional anomaly detection for indoor surveillance,'' \emph{IEEE Robotics
  and Automation Letters}, 2021.

\bibitem{haltingproblem-chi-2023}
B.~Brown, M.~Broth, and E.~Vinkhuyzen, ``The halting problem: Video analysis of
  self-driving cars in traffic,'' in \emph{CHI}, 2023.

\bibitem{neurips-2020-tinytl}
H.~Cai, C.~Gan, L.~Zhu, and S.~Han, ``Tinytl: Reduce memory, not parameters for
  efficient on-device learning,'' \emph{Advances in Neural Information
  Processing Systems}, 2020.

\bibitem{chen-eva-crop-16-bigmm}
N.~Chen, Y.~Chen, Y.~You, H.~Ling, P.~Liang, and R.~Zimmermann, ``Dynamic urban
  surveillance video stream processing using fog computing,'' in \emph{BigMM},
  2016.

\bibitem{chen-eva-reuse-15-sensys}
T.~Y.-H. Chen, L.~Ravindranath, S.~Deng, P.~Bahl, and H.~Balakrishnan,
  ``Glimpse: Continuous, real-time object recognition on mobile devices,'' in
  \emph{SenSys}, 2015.

\bibitem{aaai-2023-mobiletl}
H.-Y. Chiang, N.~Frumkin, F.~Liang, and D.~Marculescu, ``Mobiletl: On-device
  transfer learning with inverted residual blocks,'' in \emph{AAAI}, 2023.

\bibitem{courbariaux-quant-15-nips}
M.~Courbariaux, Y.~Bengio, and J.-P. David, ``Binaryconnect: Training deep
  neural networks with binary weights during propagations,'' \emph{NIPS}, 2015.

\bibitem{courbariaux-iclr-15}
M.~Courbariaux, Y.~Bengio, and J.~David, ``Low precision arithmetic for deep
  learning,'' in \emph{ICLR}, 2015.

\bibitem{dao-cl-filter-17-mass}
T.~Dao, A.~Roy-Chowdhury, N.~Nasrabadi, S.~V. Krishnamurthy, P.~Mohapatra, and
  L.~M. Kaplan, ``Accurate and timely situation awareness retrieval from a
  bandwidth constrained camera network,'' in \emph{MASS}, 2017.

\bibitem{ms-fp:darvish:nips:2020}
B.~Darvish~Rouhani, D.~Lo, R.~Zhao, M.~Liu, J.~Fowers, K.~Ovtcharov,
  A.~Vinogradsky, S.~Massengill, L.~Yang, R.~Bittner \emph{et~al.}, ``{Pushing
  the Limits of Narrow Precision Inferencing at Cloud Scale with Microsoft
  Floating Point},'' in \emph{NeurIPS}, 2020.

\bibitem{mx-isca-2023}
B.~Darvish~Rouhani, R.~Zhao, V.~Elango, R.~Shafipour, M.~Hall,
  M.~Mesmakhosroshahi, A.~More, L.~Melnick, M.~Golub, G.~Varatkar
  \emph{et~al.}, ``With shared microexponents, a little shifting goes a long
  way,'' in \emph{ISCA}, 2023.

\bibitem{www-2023-edgemove}
Z.~Dong, Q.~He, F.~Chen, H.~Jin, T.~Gu, and Y.~Yang, ``Edgemove: Pipelining
  device-edge model training for mobile intelligence,'' in \emph{Proceedings of
  the ACM Web Conference 2023}, 2023.

\bibitem{drolia-eva-reuse-17-sec}
U.~Drolia, K.~Guo, and P.~Narasimhan, ``Precog: Prefetching for image
  recognition applications at the edge,'' in \emph{SEC}, 2017.

\bibitem{drolia-eva-reuse-17-icdcs}
U.~Drolia, K.~Guo, J.~Tan, R.~Gandhi, and P.~Narasimhan, ``Cachier:
  Edge-caching for recognition applications,'' in \emph{ICDCS}, 2017.

\bibitem{hbfp:2018}
M.~Drumond, T.~LIN, M.~Jaggi, and B.~Falsafi, ``{Training DNNs with Hybrid
  Block Floating Point},'' in \emph{NeurIPS}, 2018.

\bibitem{8794073}
S.~Dutta and C.~Ekenna, ``{Air-to-Ground Surveillance Using Predictive
  Pursuit},'' in \emph{ICRA}, 2019.

\bibitem{TBP-Former}
S.~Fang, Z.~Wang, Y.~Zhong, J.~Ge, S.~Chen, and Y.~Wang, ``{TBP-Former:
  Learning Temporal Bird's-Eye-View Pyramid for Joint Perception and Prediction
  in Vision-Centric Autonomous Driving},'' in \emph{CVPR}, 2023.

\bibitem{planaria:2020}
S.~Ghodrati, B.~H. Ahn, J.~K. Kim, S.~Kinzer, B.~R. Yatham, N.~Alla, H.~Sharma,
  M.~Alian, E.~Ebrahimi, N.~S. Kim \emph{et~al.}, ``Planaria: Dynamic
  architecture fission for spatial multi-tenant acceleration of deep neural
  networks,'' in \emph{MICRO}, 2020.

\bibitem{ghunaim-cl-23-cvpr}
Y.~Ghunaim, A.~Bibi, K.~Alhamoud, M.~Alfarra, H.~A. A.~K. Hammoud, A.~Prabhu,
  P.~H. Torr, and B.~Ghanem, ``Real-time evaluation in online continual
  learning: A new hope,'' in \emph{CVPR}, 2023.

\bibitem{tpu:edge}
{Google}, ``Edge tpu,'' \url{https://cloud.google.com/edge-tpu/}, 2018.

\bibitem{bfloat16-google:2019}
Google, ``Bfloat16: The secret to high performance on cloud tpus,''
  \url{https://cloud.google.com/blog/products/ai-machine-learning/bfloat16-the-secret-to-high-performance-on-cloud-tpus?hl=en},
  2019.

\bibitem{guo-eva-reuse-18-mobicom}
P.~Guo, B.~Hu, R.~Li, and W.~Hu, ``Foggycache: Cross-device approximate
  computation reuse,'' in \emph{MobiCom}, 2018.

\bibitem{guo-eva-crop-19-tomm}
Y.~Guo, B.~Zou, J.~Ren, Q.~Liu, D.~Zhang, and Y.~Zhang, ``Distributed and
  efficient object detection via interactions among devices, edge, and cloud,''
  \emph{ToMM}, 2019.

\bibitem{hailo-8}
{Hailo}, ``Hailo expands hail-8 ai accelerator portfolio,''
  \url{https://hailo.ai/hailo-8l-entry-level-ai-accelerator-announcement/},
  2023.

\bibitem{fp16-17-hpec}
N.-M. Ho and W.-F. Wong, ``Exploiting half precision arithmetic in nvidia
  gpus,'' in \emph{HPEC}, 2017.

\bibitem{online-distillation}
J.~Houyon, A.~Cioppa, Y.~Ghunaim, M.~Alfarra, A.~Halin, M.~Henry, B.~Ghanem,
  and M.~V. Droogenbroeck, ``{Online Distillation With Continual Learning for
  Cyclic Domain Shifts},'' in \emph{CVPR}, 2023.

\bibitem{st-p3}
S.~Hu, L.~Chen, P.~Wu, H.~Li, J.~Yan, and D.~Tao, ``{ST-P3: End-to-End
  Vision-Based Autonomous Driving via Spatial-Temporal Feature Learning},'' in
  \emph{ECCV}, 2022.

\bibitem{collate}
S.~Huai, D.~Liu, H.~Kong, X.~Luo, W.~Liu, R.~Subramaniam, C.~Makaya, and
  Q.~Lin, ``{Collate: Collaborative Neural Network Learning for
  Latency-Critical Edge Systems},'' in \emph{ICCD}, 2022.

\bibitem{zerobn}
S.~Huai, L.~Zhang, D.~Liu, W.~Liu, and R.~Subramaniam, ``{ZeroBN: Learning
  Compact Neural Networks For Latency-Critical Edge Systems},'' in \emph{DAC},
  2021.

\bibitem{iccv-2023-gameformer}
Z.~Huang, H.~Liu, and C.~Lv, ``Gameformer: Game-theoretic modeling and learning
  of transformer-based interactive prediction and planning for autonomous
  driving,'' in \emph{ICCV}, 2023.

\bibitem{binarized-quant-16-nips}
I.~Hubara, M.~Courbariaux, D.~Soudry, R.~El-Yaniv, and Y.~Bengio, ``Binarized
  neural networks,'' \emph{NIPS}, 2016.

\bibitem{hung-eva-part-18-sec}
C.-C. Hung, G.~Ananthanarayanan, P.~Bodik, L.~Golubchik, M.~Yu, P.~Bahl, and
  M.~Philipose, ``Videoedge: Processing camera streams using hierarchical
  clusters,'' in \emph{SEC}, 2018.

\bibitem{iot-jia-2022}
L.~Jia, Z.~Zhou, F.~Xu, and H.~Jin, ``Cost-efficient continuous edge learning
  for artificial intelligence of things,'' \emph{IEEE Internet of Things
  Journal}, 2022.

\bibitem{cvpr-2023-jia}
X.~Jia, P.~Wu, L.~Chen, J.~Xie, C.~He, J.~Yan, and H.~Li, ``Think twice before
  driving: Towards scalable decoders for end-to-end autonomous driving,'' in
  \emph{CVPR}, 2023.

\bibitem{jung-cl-reg-20-nips}
S.~Jung, H.~Ahn, S.~Cha, and T.~Moon, ``Continual learning with node-importance
  based adaptive group sparse regularization,'' \emph{NIPS}, 2020.

\bibitem{neurosurgeon-eva-part-17-asplos}
Y.~Kang, J.~Hauswald, C.~Gao, A.~Rovinski, T.~Mudge, J.~Mars, and L.~Tang,
  ``Neurosurgeon: Collaborative intelligence between the cloud and mobile
  edge,'' in \emph{ASPLOS}, 2017.

\bibitem{emsoft-2023-dacapo}
O.~Khan, G.~Park, and E.~Seo, ``Dacapo: An on-device learning scheme for
  memory-constrained embedded systems,'' \emph{ACM Transactions on Embedded
  Computing Systems}, 2023.

\bibitem{recl-nsdi-2023}
M.~Khani, G.~Ananthanarayanan, K.~Hsieh, J.~Jiang, R.~Netravali, Y.~Shu,
  M.~Alizadeh, and V.~Bahl, ``{RECL: Responsive Resource-Efficient Continuous
  Learning for Video Analytics},'' in \emph{NSDI}, 2023.

\bibitem{kirkpatrick-cl-reg-17-pnas}
J.~Kirkpatrick, R.~Pascanu, N.~Rabinowitz, J.~Veness, G.~Desjardins, A.~A.
  Rusu, K.~Milan, J.~Quan, T.~Ramalho, A.~Grabska-Barwinska \emph{et~al.},
  ``Overcoming catastrophic forgetting in neural networks,'' \emph{PNAS}, 2017.

\bibitem{10247965}
H.~Kong, D.~Liu, X.~Luo, S.~Huai, R.~Subramaniam, C.~Makaya, Q.~Lin, and
  W.~Liu, ``{Towards Efficient Convolutional Neural Network for Embedded
  Hardware via Multi-Dimensional Pruning},'' in \emph{DAC}, 2023.

\bibitem{eomu-mm-2023}
Y.~Kong, P.~Yang, and Y.~Cheng, ``{Edge-Assisted On-Device Model Update for
  Video Analytics in Adverse Environments},'' in \emph{MM}, 2023.

\bibitem{flexpoint:2017}
U.~K{\"o}ster, T.~Webb, X.~Wang, M.~Nassar, A.~K. Bansal, W.~Constable,
  O.~Elibol, S.~Gray, S.~Hall, L.~Hornof \emph{et~al.}, ``{Flexpoint: An
  Adaptive Numerical Format for Efficient Training of Deep Neural Networks},''
  in \emph{NeurIPS}, 2017.

\bibitem{lee-cl-reg-20-cvpr}
J.~Lee, H.~G. Hong, D.~Joo, and J.~Kim, ``Continual learning with extended
  kronecker-factored approximate curvature,'' in \emph{CVPR}, 2020.

\bibitem{dataflowmirroring:2021}
J.~Lee, J.~Choi, J.~Kim, J.~Lee, and Y.~Kim, ``Dataflow mirroring:
  Architectural support for highly efficient fine-grained spatial multitasking
  on systolic-array npus,'' in \emph{DAC}, 2021.

\bibitem{carm-cl-22-dac}
S.~Lee, M.~Weerakoon, J.~Choi, M.~Zhang, D.~Wang, and M.~Jeon, ``{CarM}:
  Hierarchical episodic memory for continual learning,'' in \emph{DAC}, 2022.

\bibitem{li-eva-part-16-sec}
D.~Li, T.~Salonidis, N.~V. Desai, and M.~C. Chuah, ``Deepcham: Collaborative
  edge-mediated adaptive deep learning for mobile object recognition,'' in
  \emph{SEC}, 2016.

\bibitem{iccv-2023-li}
S.~Li, Y.~Yang, D.~Zeng, and X.~Wang, ``Adaptive and background-aware vision
  transformer for real-time uav tracking,'' in \emph{ICCV}, 2023.

\bibitem{GLTF-MA}
Y.~Li, D.~Yuan, M.~Sun, H.~Wang, X.~Liu, and J.~Liu, ``{Generalized UAV Object
  Detection via Frequency Domain Disentanglement},'' in \emph{CVPR}, 2023.

\bibitem{lin-quant-16-icml}
D.~Lin, S.~Talathi, and S.~Annapureddy, ``Fixed point quantization of deep
  convolutional networks,'' in \emph{ICML}.\hskip 1em plus 0.5em minus
  0.4em\relax PMLR, 2016, pp. 2849--2858.

\bibitem{liu-eva-part-18-edgesys}
P.~Liu, B.~Qi, and S.~Banerjee, ``Edgeeye: An edge service framework for
  real-time intelligent video analytics,'' in \emph{EdgeSys}, 2018.

\bibitem{UAVMOT}
S.~Liu, X.~Li, H.~Lu, and Y.~He, ``{Multi-Object Tracking Meets Moving UAV},''
  in \emph{CVPR}, 2022.

\bibitem{iccv-2023-aerialvln}
S.~Liu, H.~Zhang, Y.~Qi, P.~Wang, Y.~Zhang, and Q.~Wu, ``Aerialvln:
  Vision-and-language navigation for uavs,'' in \emph{ICCV}, 2023.

\bibitem{ptq-for-vit-neurips-2021}
Z.~Liu, Y.~Wang, K.~Han, W.~Zhang, S.~Ma, and W.~Gao, ``{Post-Training
  Quantization for Vision Transformer},'' in \emph{NeurIPS}, 2021.

\bibitem{mobicom-2023-miro}
X.~Ma, S.~Jeong, M.~Zhang, D.~Wang, J.~Choi, and M.~Jeon, ``Cost-effective
  on-device continual learning over memory hierarchy with miro,'' in
  \emph{Proceedings of the 29th Annual International Conference on Mobile
  Computing and Networking}, 2023.

\bibitem{micikevicius:mixed-precision:iclr:2018}
P.~Micikevicius, S.~Narang, J.~Alben, G.~Diamos, E.~Elsen, D.~Garcia,
  B.~Ginsburg, M.~Houston, O.~Kuchaiev, G.~Venkatesh \emph{et~al.}, ``{Mixed
  Precision Training},'' in \emph{ICLR}, 2018.

\bibitem{cvpr-2023-mobilevos}
R.~Miles, M.~K. Yucel, B.~Manganelli, and A.~Sa{\`a}-Garriga, ``Mobilevos:
  Real-time video object segmentation contrastive learning meets knowledge
  distillation,'' in \emph{CVPR}, 2023.

\bibitem{hpca-2024-usas}
C.~S. Mishra, J.~Sampson, M.~T. Kandemir, V.~Narayanan, and C.~R. Das, ``Usas:
  A sustainable continuous-learning framework for edge servers,'' in
  \emph{HPCA}, 2024.

\bibitem{omd}
R.~T. Mullapudi, S.~Chen, K.~Zhang, D.~Ramanan, and K.~Fatahalian, ``{Online
  Model Distillation for Efficient Video Inference},'' in \emph{ICCV}, 2019.

\bibitem{flex-block:2022}
S.-H. Noh, J.~Koo, S.~Lee, J.~Park, and J.~Kung, ``{FlexBlock: A Flexible DNN
  Training Accelerator with Multi-Mode Block Floating Point Support},'' in
  \emph{arXiv}, 2022.

\bibitem{nvidia-tf32}
{NVIDIA}, ``Getting immediate speedups with nvidia a100 tf32,''
  \url{https://developer.nvidia.com/blog/getting-immediate-speedups-with-a100-tf32/},
  2020.

\bibitem{nvidia-jetson}
{NVIDIA}, ``Nvidia jetson,''
  \url{https://www.nvidia.com/en-us/autonomous-machines/embedded-systems/},
  2023.

\bibitem{paik-cl-reg-20-aaai}
I.~Paik, S.~Oh, T.~Kwak, and I.~Kim, ``Overcoming catastrophic forgetting by
  neuron-level plasticity control,'' in \emph{AAAI}, 2020.

\bibitem{icra-2023-pichierri}
L.~Pichierri, G.~Carnevale, L.~Sforni, A.~Testa, and G.~Notarstefano, ``A
  distributed online optimization strategy for cooperative robotic
  surveillance,'' in \emph{ICRA}, 2023.

\bibitem{prabhu-cl-23-cvpr}
A.~Prabhu, H.~A. A.~K. Hammoud, P.~Dokania, P.~H. Torr, S.-N. Lim, B.~Ghanem,
  and A.~Bibi, ``Computationally budgeted continual learning: What does
  matter?'' in \emph{CVPR}, 2023.

\bibitem{fast:2022}
S.~Qian~Zhang, B.~McDanel, and H.~T. Kung, ``{FAST: DNN Training Under Variable
  Precision Block Floating Point with Stochastic Rounding},'' in \emph{HPCA},
  2022.

\bibitem{sigma:2020}
E.~Qin, A.~Samajdar, H.~Kwon, V.~Nadella, S.~Srinivasan, D.~Das, B.~Kaul, and
  T.~Krishna, ``{SIGMA: A Sparse and Irregular GEMM Accelerator with Flexible
  Interconnects for DNN Training},'' in \emph{HPCA}, 2020.

\bibitem{binary-tenary-quant-21-cvpr}
R.~Razani, G.~Morin, E.~Sari, and V.~P. Nia, ``Adaptive binary-ternary
  quantization,'' in \emph{CVPR}, 2021.

\bibitem{icarl-cl-17-cvpr}
S.-A. Rebuffi, A.~Kolesnikov, G.~Sperl, and C.~H. Lampert, ``icarl: Incremental
  classifier and representation learning,'' in \emph{CVPR}, 2017.

\bibitem{scalesim}
A.~Samajdar, Y.~Zhu, P.~Whatmough, M.~Mattina, and T.~Krishna, ``Scale-sim:
  Systolic cnn accelerator simulator,'' \emph{arXiv preprint arXiv:1811.02883},
  2018.

\bibitem{sarwar-cl-reg-19-paml}
S.~S. Sarwar, A.~Ankit, and K.~Roy, ``Incremental learning in deep
  convolutional neural networks using partial network sharing,'' \emph{PAML},
  2019.

\bibitem{ral-2022-seo}
M.~Seo, D.~Cho, S.~Lee, J.~Park, D.~Kim, J.~Lee, J.~Ju, H.~Noh, and D.-G. Choi,
  ``A self-supervised sampler for efficient action recognition: Real-world
  applications in surveillance systems,'' \emph{IEEE Robotics and Automation
  Letters}, 2022.

\bibitem{bitfusion:isca18}
H.~Sharma, J.~Park, N.~Suda, L.~Lai, B.~Chau, V.~Chandra, and H.~Esmaeilzadeh,
  ``Bit fusion: Bit-level dynamically composable architecture for accelerating
  deep neural networks,'' in \emph{ISCA}, 2018.

\bibitem{shim-cl-21-aaai}
D.~Shim, Z.~Mai, J.~Jeong, S.~Sanner, H.~Kim, and J.~Jang, ``Online
  class-incremental continual learning with adversarial shapley value,'' in
  \emph{AAAI}, 2021.

\bibitem{ams}
M.~K. Shirkoohi, P.~Hamadanian, A.~Nasr{-}Esfahany, and M.~Alizadeh,
  ``{Real-Time Video Inference on Edge Devices via Adaptive Model Streaming},''
  in \emph{CVPR}, 2021.

\bibitem{arm-ethos}
A.~Skillman and T.~Edso, ``A technical overview of cortex-m55 and ethos-u55:
  Arm’s most capable processors for endpoint ai,'' in \emph{HotChips}, 2020.

\bibitem{fp8-19-nips}
X.~Sun, J.~Choi, C.-Y. Chen, N.~Wang, S.~Venkataramani, V.~V. Srinivasan,
  X.~Cui, W.~Zhang, and K.~Gopalakrishnan, ``Hybrid 8-bit floating point (hfp8)
  training and inference for deep neural networks,'' \emph{NIPS}, 2019.

\bibitem{odin}
A.~Suprem, J.~Arulraj, C.~Pu, and J.~E. Ferreira, ``{ODIN: Automated Drift
  Detection and Recovery in Video Analytics},'' in \emph{VLDB}, 2020.

\bibitem{tesla-20-micro}
E.~Talpes, D.~D. Sarma, G.~Venkataramanan, P.~Bannon, B.~McGee, B.~Floering,
  A.~Jalote, C.~Hsiong, S.~Arora, A.~Gorti \emph{et~al.}, ``Compute solution
  for tesla's full self-driving computer,'' \emph{MICRO}, 2020.

\bibitem{Tijtgat}
N.~Tijtgat, W.~Van~Ranst, T.~Goedeme, B.~Volckaert, and F.~De~Turck,
  ``{Embedded Real-Time Object Detection for a UAV Warning System},'' in
  \emph{ICCV}, 2017.

\bibitem{wang-cl-filter-18-sec}
J.~Wang, Z.~Feng, Z.~Chen, S.~George, M.~Bala, P.~Pillai, S.-W. Yang, and
  M.~Satyanarayanan, ``Bandwidth-efficient live video analytics for drones via
  edge computing,'' in \emph{SEC}, 2018.

\bibitem{dac-2023-shoggoth}
L.~Wang, K.~Lu, N.~Zhang, X.~Qu, J.~Wang, J.~Wan, G.~Li, and J.~Xiao,
  ``Shoggoth: Towards efficient edge-cloud collaborative real-time video
  inference via adaptive online learning,'' in \emph{DAC}, 2023.

\bibitem{afec-cl-reg-21-nips}
L.~Wang, M.~Zhang, Z.~Jia, Q.~Li, C.~Bao, K.~Ma, J.~Zhu, and Y.~Zhong, ``Afec:
  Active forgetting of negative transfer in continual learning,'' \emph{NIPS},
  2021.

\bibitem{fp8-18-nips}
N.~Wang, J.~Choi, D.~Brand, C.-Y. Chen, and K.~Gopalakrishnan, ``Training deep
  neural networks with 8-bit floating point numbers,'' \emph{NIPS}, 2018.

\bibitem{icra-2021-pheromone}
T.~Wang, G.~Dong, and P.~Huang, ``Pheromone-diffusion-based conscientious
  reactive path planning for road network persistent surveillance,'' in
  \emph{ICRA}, 2021.

\bibitem{cvpr-2023-xiong}
X.~Xiong, Y.~Liu, T.~Yuan, Y.~Wang, Y.~Wang, and H.~Zhao, ``Neural map prior
  for autonomous driving,'' in \emph{CVPR}, 2023.

\bibitem{cvpr-2022-rep}
L.~Yang, A.~S. Rakin, and D.~Fan, ``Repnet: Efficient on-device learning via
  feature reprogramming,'' in \emph{CVPR}, 2022.

\bibitem{lavea-eva-part-17-sec}
S.~Yi, Z.~Hao, Q.~Zhang, Q.~Zhang, W.~Shi, and Q.~Li, ``Lavea: Latency-aware
  video analytics on edge computing platform,'' in \emph{SEC}, 2017.

\bibitem{yoon-cl-22-iclr}
J.~Yoon, D.~Madaan, E.~Yang, and S.~J. Hwang, ``Online coreset selection for
  rehearsal-based continual learning,'' in \emph{ICLR}, 2022.

\bibitem{bdd100k}
F.~Yu, H.~Chen, X.~Wang, W.~Xian, Y.~Chen, F.~Liu, V.~Madhavan, and T.~Darrell,
  ``Bdd100k: A diverse driving dataset for heterogeneous multitask learning,''
  in \emph{Proceedings of the IEEE/CVF conference on computer vision and
  pattern recognition}, 2020, pp. 2636--2645.

\bibitem{ptq4vit-eccv-2022}
Z.~Yuan, C.~Xue, Y.~Chen, Q.~Wu, and G.~Sun, ``{PTQ4ViT: Post-training
  Quantization for Vision Transformers with Twin Uniform Quantization},'' in
  \emph{ECCV}, 2022.

\bibitem{ffsva-cl-filter-18-icpp}
C.~Zhang, Q.~Cao, H.~Jiang, W.~Zhang, J.~Li, and J.~Yao, ``Ffs-va: A fast
  filtering system for large-scale video analytics,'' in \emph{ICPP}, 2018.

\bibitem{active-cl-sensys-2022}
L.~Zhang, G.~Gao, and H.~Zhang, ``{Towards Data-Efficient Continuous Learning
  for Edge Video Analytics via Smart Caching},'' in \emph{SenSys}, 2022.

\bibitem{VTUAV}
P.~Zhang, J.~Zhao, D.~Wang, H.~Lu, and X.~Ruan, ``{Visible-Thermal UAV
  Tracking: A Large-Scale Benchmark and New Baseline},'' in \emph{CVPR}, 2022.

\bibitem{hpca-2024-camel}
S.~Q. Zhang, T.~Tambe, N.~Cuevas, G.-Y. Wei, and D.~Brooks, ``Camel:
  Co-designing ai models and edrams for efficient on-device learning,'' in
  \emph{HPCA}, 2024.

\bibitem{dorefa}
S.~Zhou, Z.~Ni, X.~Zhou, H.~Wen, Y.~Wu, and Y.~Zou, ``Dorefa-net: Training low
  bitwidth convolutional neural networks with low bitwidth gradients,''
  \emph{CoRR}, 2016.

\bibitem{tenary-quant-17-iclr}
C.~Zhu, S.~Han, H.~Mao, and W.~J. Dally, ``Trained ternary quantization,'' in
  \emph{ICLR}, 2017.

\end{thebibliography}

\end{document}